\documentclass[12pt,a4paper]{article}
\usepackage{amsmath}
\usepackage{amsfonts}
\usepackage{graphicx}
\usepackage{psfrag}

\begin{document}
\numberwithin{equation}{section}

\rightline{DCPT-05/05}   
\rightline{hep-th/0502195}   
\vfil


\begin{center} 
{\Large \bf Black hole thermodynamics}
\end{center} 
\vskip 1cm   
  
\renewcommand{\thefootnote}{\fnsymbol{footnote}}   
\centerline{\bf Simon 
F. Ross\footnote{S.F.Ross@durham.ac.uk}}     
\vskip .5cm   
\centerline{ \it Centre for Particle Theory, Department of  
Mathematical Sciences}   
\centerline{\it University of Durham, South Road, Durham DH1 3LE, U.K.}   
  
\setcounter{footnote}{0}   
\renewcommand{\thefootnote}{\arabic{footnote}}


\begin{abstract}  
These notes introduce basic aspects of black hole thermodynamics. I
review the classical laws of black hole mechanics, give a brief
introduction to the essential concepts of quantum field theory in
curved spacetime, and derive the Unruh and Hawking effects. I conclude
with a discussion of entropy from the Euclidean path integral point
of view and in the context of the AdS/CFT correspondence. Originally
delivered as a set of lectures at the PIMS summer school ``Strings,
Gravity and Cosmology'' in August 2004. 
\end{abstract}

\titlepage

\tableofcontents

\section{Introduction}
\label{intro}

The aim of these notes is to provide a brief introduction to the basic
ideas of black hole thermodynamics. The principal aim is to explain
the association of a temperature and entropy with stationary black
holes,
\begin{equation} \label{bhthermo}
T = \frac{ \hbar \kappa}{2 \pi}, \quad S = \frac{A}{4 G \hbar},
\end{equation}
where $\kappa$ is the black hole's surface gravity and $A$ is the area
of the event horizon. This is not intended as a review article; this
is a large subject, and there are aspects I am not expert enough to
review. It will reflect my own perspective, filtered through the
desire to present a hopefully useful pedagogical introduction. I will
focus on providing an explanation of the key results in the simplest
possible contexts. Two points I intend to highlight are the
universality of the results---the relations \eqref{bhthermo} apply to
any stationary black hole\footnote{This is for Einstein's general
relativity. To extend it to theories with higher-curvature terms in
the Lagrangian, the expression for the entropy in \eqref{bhthermo} is
replaced by a more general integral over the event horizon, as
described in section \ref{first}.}---and the central role of
regularity at the event horizon in the derivation of the thermodynamic
properties. I will emphasise the Euclidean point of view on black hole
thermodynamics. 

Other useful introductory reviews
are~\cite{jacthermo,waldthermo1,waldthermo2,jennie,padmanabhan,pagethermo}. An
in-depth discussion of the derivation of Hawking radiation is given
in~\cite{primer}, and the connection between Killing horizons and
thermal properties is discussed at length in~\cite{fullr}. Most
monographs on quantum field theory in curved spacetime include some
discussion of black hole thermodynamics (see the list of references
later in the introduction): a particularly comprehensive discussion is
given in~\cite{wald94}. The recent textbook~\cite{carroll} also has a
nice treatment of the subject. For detailed discussion of black holes
and string theory, see~\cite{peet:tasi,david}.

Black hole thermodynamics has continued to fascinate researchers over
the 30 years since Hawking's discovery of the thermal radiation from
black holes~\cite{hrad1,hrad2} because it provides a real connection
between gravity and quantum mechanics. The relation between
geometrical properties of the event horizon and thermodynamic
quantities provides a clear indication that there is a relation
between properties of the spacetime geometry and some kind of quantum
physics. The thermodynamic behaviour \eqref{bhthermo} should have a
statistical interpretation in quantum gravity, and hence it provides
clues to the form of the quantisation of gravity. In string theory in
particular, the scaling of $S$ with the area was argued to imply that
the fundamental theory cannot have local bulk degrees of freedom in
the same way as an ordinary quantum field
theory~\cite{thooft,susskind,holrev}, and the understanding of these
relations in a string theory context and providing them with a natural
statistical explanation has been one of the driving forces in the
revolution which has taken us beyond the formulation of string theory
in terms of string scattering in a fixed background spacetime to a
fully non-perturbative formulation of the theory through the AdS/CFT
dualities (see for example the review~\cite{agmoo}).

The study of black hole thermodynamics also played an important role
in the development of quantum field theory in curved spacetime. This
is a subject of considerable physical importance in its own right,
playing an important role in, for example, the very early universe
(see the lectures by Robert Brandenburger at this school, or the notes
online from another set of lectures~\cite{brandenberger}). The study
of such curved backgrounds has also expanded our understanding of
quantum field theory in general.

These notes begin with a review of the relevant features of the
classical black hole geometries, stressing the distinction between
Killing and event horizons, in section~\ref{rev}. I describe the
evidence for an analogy between black hole dynamics and thermodynamics
at the classical level in section~\ref{class}. Although this is the
oldest part of the subject, new developments are still occurring.  I
give a basic introduction to quantum field theory in curved spacetime
in section~\ref{qft}, introducing the technology required to
understand the argument for black hole thermodynamics. This is not
meant as a complete introduction to this rich subject, and the
interested reader would be well-advised to consult a more extensive
discussion: there are several excellent
texts~\cite{birdav,haag,fullingbook,wald94,dewitt03} and reviews
(e.g., ~\cite{dewitt75,gibbons79,dewitt79,ford,jac}) which cover it; I
particularly recommend the recent review by Jacobson~\cite{jac}, which
provides a highly-accessible introduction to the modern viewpoint on
this subject.  In section~\ref{hrad}, I will discuss the Unruh effect
and Hawking radiation, showing how quantum fields on black hole
backgrounds behave thermally. Finally, in section~\ref{euclid}, the
lectures finish with a more sketchy discussion of entropy from the
Euclidean path-integral point of view and the understanding of black
hole thermodynamics in AdS/CFT. Mainly because of time constraints, I
have focused on equilibrium and quasi-equilibrium processes, and
excluded any detailed discussion of dynamical issues such as the
generalised second law.

I have attempted to make these notes reasonably self-contained and
pedagogical. The choices of topics and emphasis naturally reflects my
own interests; it was also influenced by the original audience. I have
drawn heavily on the referenced sources in preparing these notes, and
make no claim of originality for any of the content. I have attempted
to give useful references to more detailed reviews and to the original
literature; while it is impossible to give full references to all the
people who have contributed to this subject, I welcome comments and
corrections. 

\section{Review of black holes}
\label{rev}

Black hole thermodynamics involves many powerful general statements.
However, to keep the discussion as concrete as possible, I will
introduce the subject by focusing on the simplest possible examples.
The simplest black hole solution is the Schwarzschild solution. This
is a vacuum solution in general relativity in four spacetime
dimensions, for which the metric is
\begin{equation} \label{schw}
ds^2 = - \left( 1 - \frac{2M}{r} \right) dt^2 + \left( 1 -
\frac{2M}{r} \right)^{-1} dr^2 + r^2 ( d\theta^2 + \sin^2 \theta
d\phi^2). 
\end{equation}
This metric appears to have a singularity at $r=2M$. However, this is
only a coordinate singularity. We can see that the light cones in
these coordinates are closing up as we approach $r=2M$, so we can
construct a better coordinate system in that region by following the
causal structure: define new coordinates
\begin{equation}
u ,v = t \pm r_* = t \pm \left[ r + 2M \ln \left( \frac{r}{2M} -1
  \right) \right] ,
\end{equation}
so
\begin{equation}
\dot{u}, \dot{v} = \dot{t} \pm \frac{\dot{r}}{(1-2M/r)} .
\end{equation}
Thus, ingoing null rays have $u=$constant, while outgoing null rays
have $v=$constant. If we write the metric in coordinates
$(u,r,\theta,\phi)$, we can extend it across $r=2M$ along ingoing null
rays. Similarly, the metric in coordinates $(v,r,\theta,\phi)$ can be
extended across $r=2M$ along outgoing null rays. 

To cover the whole spacetime including the whole region behind the
horizon, we need to introduce the Kruskal coordinates
\begin{equation} \label{krusu}
u' = e^{u/4M} = \left( \frac{r}{2M} -1 \right)^{1/2} e^{(r+t)/4M},
\end{equation}
\begin{equation} \label{krusv}
v' = -e^{-v/4M} = \left( \frac{r}{2M} -1 \right)^{1/2} e^{(r-t)/4M}.
\end{equation}
In terms of these coordinates,
\begin{equation}
ds^2 = -\frac{32M^3}{r} e^{-r/2M} du' dv' + r^2 d\Omega_2,
\end{equation}
where $r(u,v)$ is defined implicitly by
(\ref{krusu},\ref{krusv}). These coordinates are maximal---all
geodesics either extend to infinite affine parameter without leaving
this chart or meet the singularity at $r=0$.  The singular surface at
$r=2M$ in the previous coordinates maps to $u'v' = 0$, which is
manifestly non-singular. On the other hand, $r=0$, which maps to
$u'v'=-1$, is still singular: this is a curvature singularity. More
generally, surfaces of constant $t$ are at $u'/v'$=constant, while
surfaces of constant $r$ are at $u'v'=$constant. We will later need
the form of the time-translation symmetry in these coordinates,
\begin{equation} \label{killv}
\partial_t = \frac{1}{ 4M} (u' \partial_{u'} - v' \partial_{v'}). 
\end{equation}

\begin{figure}[htbp]
\centering
\psfrag{u}{$u'$}
\psfrag{v}{$v'$}
\psfrag{r=0}{$r=0$}
\psfrag{h+}{$r=2M, t=\infty$}
\psfrag{h-}{$r=2M, t=-\infty$}
\psfrag{rconst}{$r=$constant}
\psfrag{tconst}{$t=$constant}
\includegraphics[width=0.7\textwidth]{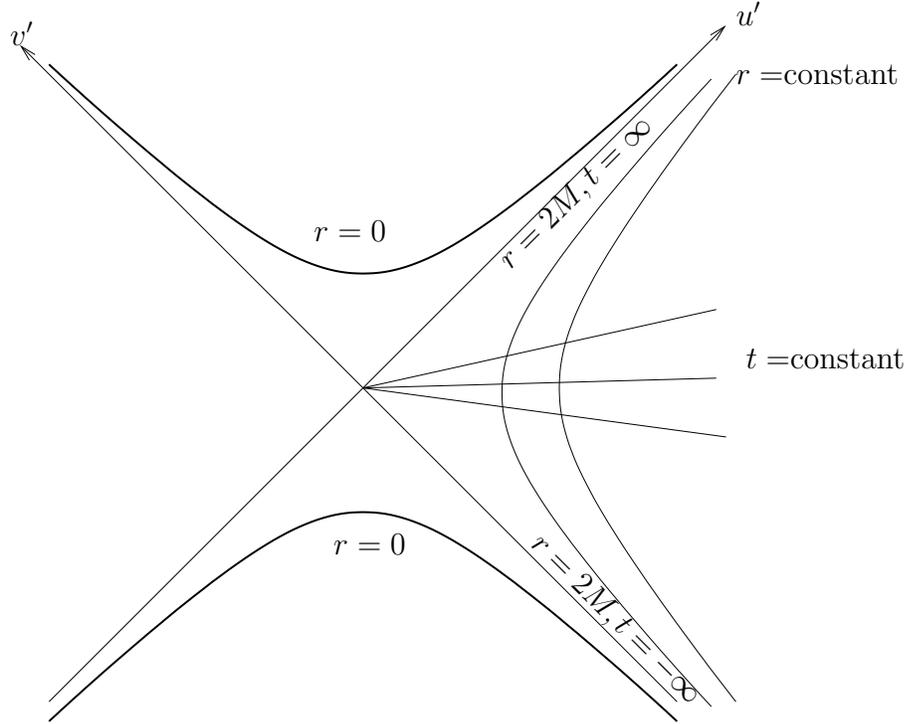}
\caption{Kruskal diagram for Schwarzschild.}
\label{fig1}
\end{figure}

These coordinates are depicted in figure~\ref{fig1}. One of the
primary advantages of this coordinate system is that the light cones
lie at 45 degrees throughout the figure. This allows us to immediately
see that $r=2M$ plays a special role. Any observer who enters the
region $r<2M$ will inevitably encounter the singularity at $r=0$.

\begin{figure}[htbp]
\centering
\psfrag{r=0}{$r=0$}
\psfrag{i+}{$\mathcal{I}^+$}
\psfrag{i-}{$\mathcal{I}^-$}
\includegraphics{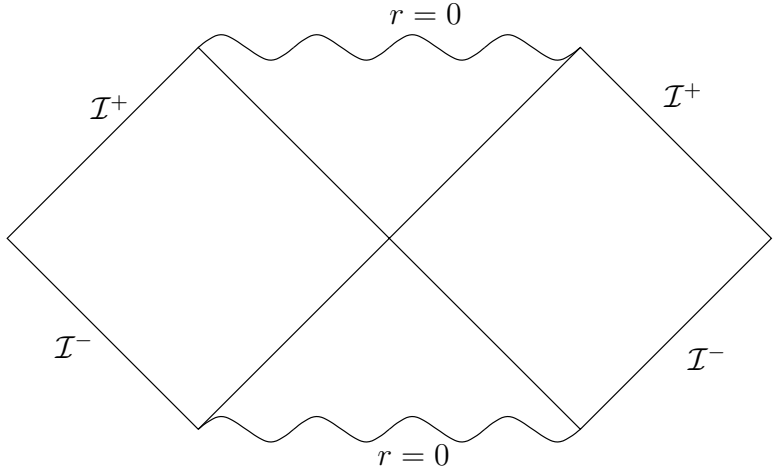}
\caption{Penrose diagram for Schwarzschild.}
\label{fig2}
\end{figure}

Another useful representation of the spacetime is the Penrose
diagram~\cite{penconf,pendiag}, which maps the whole spacetime to a
finite region. Formally, we consider a manifold $(\bar{M},\bar{g})$
such that the original spacetime $(M,g)$ maps to a subset of
$\bar{M}$, with a conformal relation between the metrics,
$\bar{g}_{\mu\nu} = \Omega^2 g_{\mu\nu}$, on the image of $M$ in
$\bar{M}$. The boundary of the image of $M$ in $\bar{M}$ is thought of
as representing the `points at infinity' in spacetime. This
construction provides a highly useful tool for discussing the global
structure of spacetimes. We can create an appropriate coordinate
system describing $\bar{M}$ by setting
\begin{equation}
\tan U = u', \quad \tan V = v'.
\end{equation}
This is represented in figure~\ref{fig2}. Once again, light cones lie
at 45 degrees in the figure. 

This maximally-extended solution provides the model and template for
our discussion of the properties of black holes and the relation to
thermodynamics. We should note, however, that for a physical black
hole formed by gravitational collapse, only a part of the spacetime is
relevant, since this vacuum solution is only applicable outside of any
collapsing matter. Furthermore, once we take Hawking radiation into
account, a physical black hole will evaporate; the quantum-corrected
black hole does not approach an equilibrium described by this
solution. Thus, this eternal Schwarzschild solution is not directly
related to the behaviour of real black holes. 

\begin{figure}[htbp]
\centering
\psfrag{r=0}{$r=0$}
\psfrag{h+}{$r=r_+, t=\infty$}
\psfrag{h-}{$r=r_+, t=-\infty$}
\psfrag{rconst}{$r=$constant}
\psfrag{tconst}{$t=$constant}
\psfrag{rinfty}{$r=\infty$}
\psfrag{i}{$\mathcal{I}$}
\includegraphics[width=0.4\textwidth]{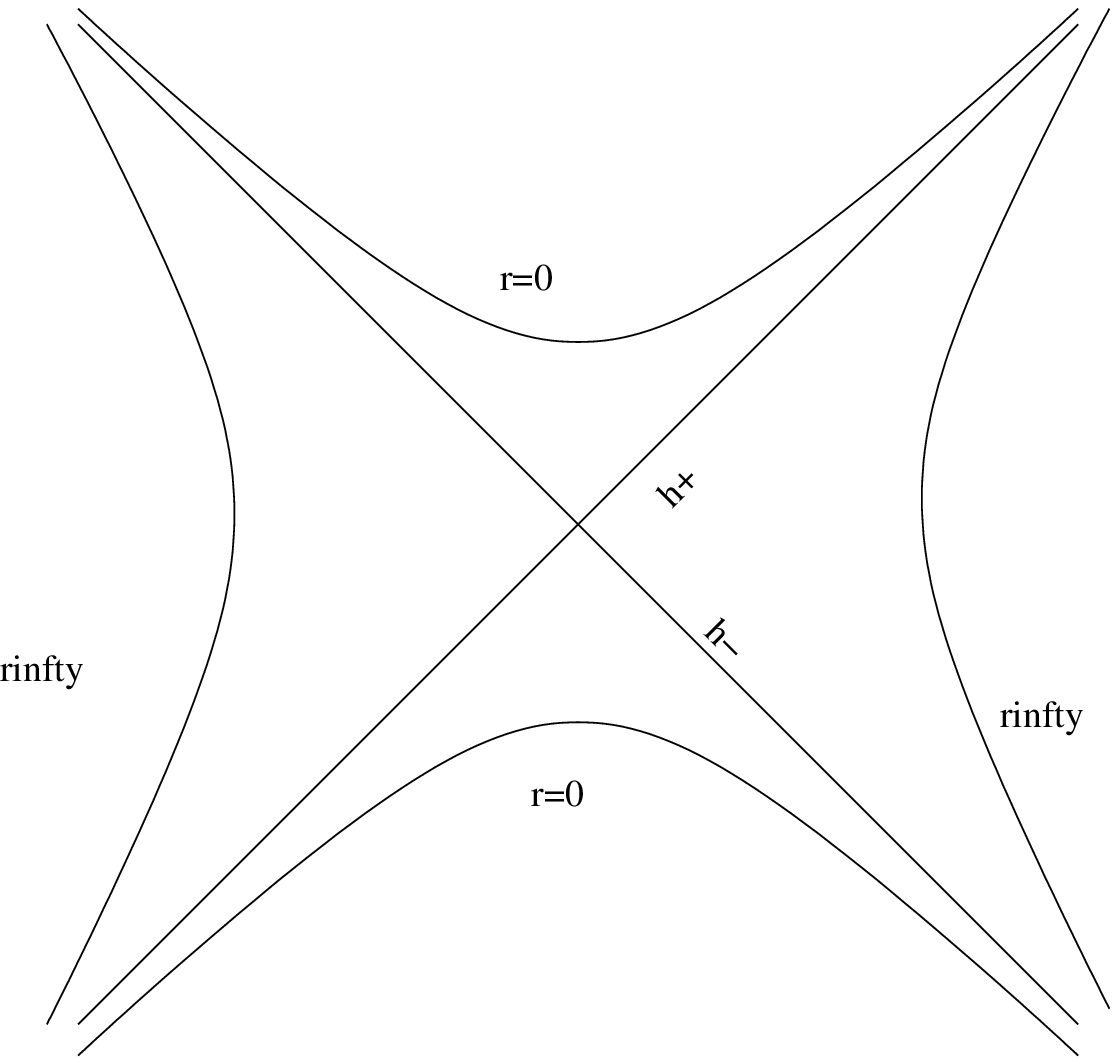}%
\hspace{0.1\textwidth}
\includegraphics[width=0.4\textwidth]{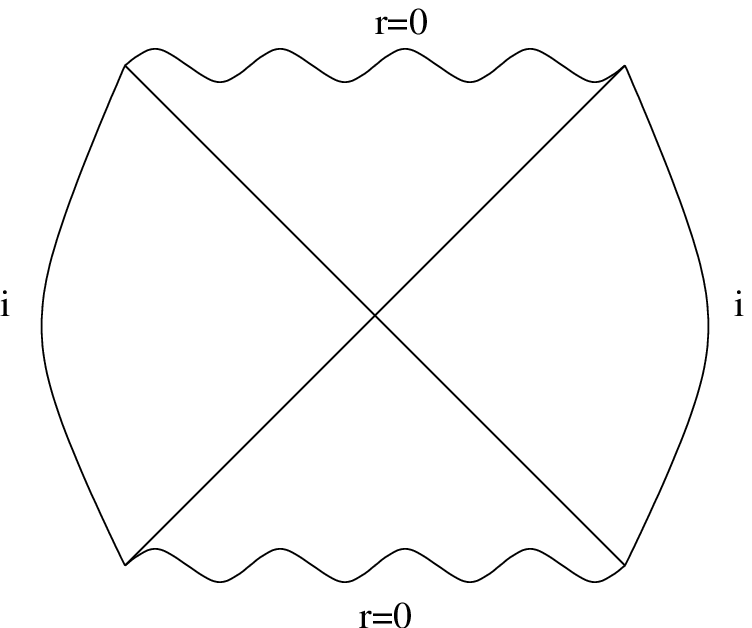}
\caption{Kruskal and Penrose diagrams for Schwarzschild-AdS.}
\label{fig3}
\end{figure}

This situation is somewhat improved if we consider a simple
generalisation, to include a negative cosmological constant. Since
this generalisation is of particular interest in string theory, we
consider the solution in $d$ spacetime dimensions, the
Schwarzschild-AdS metric
\begin{equation} \label{sads}
ds^2 = - f(r)
dt^2 + f(r)^{-1} dr^2 + r^2 d\Omega_{d-2},
\end{equation}
where 
\begin{equation} \label{fr}
f(r) = \left( 1 - \frac{w_d M}{r^{d-3}} + \frac{r^2}{\ell^2} \right),
\end{equation}
\begin{equation}
w_d = \frac{16 \pi G}{(d-2) \mbox{Vol}(S^{d-2})},
\end{equation}
$d\Omega_{d-2}$ is the metric on the unit $d-2$ sphere, and the
cosmological constant is $\Lambda = -(d-1)/\ell^2$. This has a
coordinate singularity at $r=r_+$, where $f(r_+)=0$, which can be
removed, as before, by passing to Kruskal coordinates. We define the
light cone coordinates
\begin{equation}
u,v = t \pm r_*,
\end{equation}
where $dr_* = dr/f(r)$, and then define Kruskal coordinates by
\begin{equation}
u' = e^{\kappa u}, \quad v' = - e^{-\kappa v},
\end{equation}
where
\begin{equation}
\kappa = \frac{1}{2} f'(r_+) = \frac{(d-1) r_+^2 + (d-3) \ell^2}{2r_+
    \ell^2}.
\end{equation}
The time-translation Killing vector is given by
\begin{equation} \label{adskillv}
\partial_t = \kappa (u' \partial_{u'} - v' \partial_{v'}). 
\end{equation}
In this case, $r_* \sim r^{-1}$ for large $r$, so $r_*$ has a finite
range. This implies that the Kruskal diagram will have an additional
boundary for some value of $u'v'$, as depicted in figure~\ref{fig3},
corresponding to the timelike boundary in the asymptotically AdS
spacetime.

In the Penrose diagram, this implies that the null asymptopia of the
Schwarzschild solution is replaced by a timelike $\mathcal{I}$. It was
recently noted~\cite{adspen} that for $d>3$, there is no choice of
conformal factor such that both the singularity $r=0$ and the
asymptotic boundaries $\mathcal{I}$ can be represented as straight
lines on the Penrose diagram.  

The study of the eternal black hole solution is a little more
well-motivated in this case. The negative cosmological constant will
prevent the radiation emitted from the black hole escaping to
infinity. Furthermore, the specific heat is positive for large enough
black holes (as we will see in section~\ref{hrad}), so they can reach
equilibrium with their Hawking radiation and settle down to a
configuration approximately described by this eternal black hole
solution.

The simple form of the above metrics is indicative of a genuine
simplicity in the physics of black holes, and not just the special
cases we have chosen to study. There are `no-hair' theorems which
state that Schwarzschild is the most general asymptotically flat
static vacuum solution; this was first proved for the four-dimensional
case in~\cite{vacnohair}, and for higher dimensions
in~\cite{nohairhigher}. In four dimensions, this generalises to
stationary solutions: the most general stationary asymptotically flat
black hole in four-dimensional vacuum general relativity is described
by the Kerr solution, which has just two parameters: the mass $M$ and
angular momentum $J$ (see~\cite{heusler} for a review of no-hair
theorems). The most remarkable recent discovery in black hole physics
is that no such no-hair theorem exists for stationary solutions in
$d>4$: the existence of black ring solutions~\cite{er} shows that
there can be more than one solution with a given $M,J$.
 
\subsection{Horizons}
\label{hors}

In the Schwarzschild-AdS solution, the surface at $r=r_+$ is referred
to as the black hole horizon or event horizon. It is actually an
example of two different kinds of horizons. First, the surface $v' =0$
is an event horizon for the right asymptotic region (and similarly
$u'=0$ is an event horizon for the left asymptotic region). An event
horizon corresponds to the intuitive idea of a black hole as a region
nothing can escape from: anything that crosses $v'=0$ can never
return. We can similarly define an event horizon in any spacetime with
specified asymptotic behaviour: the event horizon $\mathcal{H}$ is the
boundary of the causal past of the asymptotic infinity, $\mathcal{H} =
\dot{J}^-(\mathcal{I}^+)$ (see~\cite{hawkellis} for a more detailed
discussion).  In a general spacetime, the region inside the event
horizon which no signal can escape from, that is, the region $M -
J^-(\mathcal{I}^+)$ in a spacetime $M$, will be referred to as the black
hole. By virtue of the definition, the event horizon in an arbitrary
spacetime is always a null hypersurface. Generators may enter the
horizon, but cannot leave it.

Because Schwarzschild-AdS is a static solution, the event horizon is
also a Killing horizon. More accurately, the whole surface $r=r_+$,
consisting of the two branches $u'=0$ and $v'=0$, forms a bifurcate
Killing horizon. A Killing horizon is a null surface $\mathcal N$ whose
generators are orbits of an isometry; that is, with a Killing vector
field $\xi$ whose orbits generate $\mathcal N$. In Schwarzschild-AdS, both
$u'=0$ and $v'=0$ are Killing horizons with respect to the
time-translation Killing vector $\partial_t$.

A bifurcate Killing horizon is a structure like the space $r=r_+$: a
pair of Killing horizons of the same Killing vector which intersect
over a spacelike two-surface, called the bifurcation surface. The
Killing vector $\xi$ necessarily vanishes on the bifurcation surface;
that is, it is a collection of fixed points of the
isometry. Conversely, if a Killing vector vanishes on a spacelike
two-surface, there is a Killing horizon with that two-surface as a
bifurcation surface~\cite{wald94}. Thus the bifurcate Killing horizon
structure is entirely determined by these fixed points of the
isometry. In Schwarzschild-AdS, the set $u'v'=0$ forms a bifurcate
Killing horizon, and $u'=v'=0$ is the bifurcation two-sphere.

On a Killing horizon, the surface gravity $\kappa$ is defined by
\begin{equation}
\xi^\mu \nabla_\mu \xi^\nu = \kappa \xi^\nu \quad \mbox{on } {\mathcal N}.
\end{equation}
Equivalently, 
\begin{equation}
\kappa^2 = -\frac{1}{2} (\nabla^\mu \xi^\nu) (\nabla_\mu \xi_\nu)
\quad \mbox{on } {\mathcal N}. 
\end{equation}
In Schwarzschild, we can work out from \eqref{killv} that $\kappa =
\pm 1/4M$.  Similarly, in the Schwarzschild-AdS solution, $r=r_+$ is a
bifurcate Killing horizon with Killing vector $\xi = \partial_t$, and
$\kappa = f'(r_+)/2$ is the surface gravity of this Killing horizon.

In general dynamical contexts, these two types of horizons are quite
different. However, for the stationary black holes, they tend to
coincide. That is, if the black hole solution has a Killing vector
which is timelike at infinity, then in many cases of interest, the
event horizon will also be a Killing horizon. This has been proved
assuming that the solution is vacuum or electrovac~\cite{carter}, or
if it has an additional angular symmetry, so that the black hole is
stationary-axisymmetric~\cite{hawkellis}. 

The notion of an event horizon plays the central role in the general
classical definition of a black hole, and in dynamical considerations
such as the generalised second law. The Killing horizon structure will
play the central role in the discussion of quantum field theory on
this spacetime, and the derivation of black hole thermodynamics in
section~\ref{hrad} will be based on the fact that the black hole
horizon is a bifurcate Killing horizon. Thus, the Killing horizon
structure is the more important for the purposes of these
lectures. Note also that Killing horizons (unlike event horizons) do
not occur only in black holes; there are many examples of Killing
horizons which are not also black hole horizons. In fact, as we see
next, this structure also occurs in flat space.

\subsection{Near-horizon structure}
\label{rindler}

We have argued above that the horizons in stationary black hole
spacetimes will typically be bifurcate Killing horizons. Many of the
universal features of black hole thermodynamics find their origin in
this common structure, so we will now study the region of the
spacetime near the horizon more closely. We focus on a neighbourhood
of a null generator of the horizon in the Schwarzschild-AdS solution
(the one at $\theta =0$, say) by considering
\begin{equation}
r = r_+ + \epsilon^2 \frac{f'(r_+)}{4} x^2,
\end{equation}
for some small $\epsilon$, so $f(r)^{-1} dr^2 = \epsilon^2 dx^2$ up to
corrections of order $\epsilon^4$, and also focus on small values of
$\theta$, $\theta = \epsilon \rho$. In this limit the metric becomes
\begin{equation} \label{n1}
ds^2 = \epsilon^2 \left( -\frac{f'(r_+)^2}{4} x^2 dt^2 + dx^2 +
ds^2_{\mathbb{R}^{(d-2)}} \right) + O(\epsilon^4).
\end{equation}
The leading part of the metric for small $\epsilon$ describes flat
space (as we might expect, since we have focused on a small part of
the spacetime), but not in the usual Cartesian coordinates. This
coordinate system is called Rindler coordinates. If we set
\begin{equation} \label{rflat}
X = x \cosh (\kappa t), \quad T = x \sinh (\kappa t),
\end{equation}
where $\kappa = f'(r_+)/2$ as before, we recover flat space in the
usual Cartesian coordinates,
\begin{equation}
ds^2 = \epsilon^2 (-dT^2 + dX^2 + ds^2_{\mathbb{R}^{(d-2)}}).
\end{equation}
The Rindler coordinates of \eqref{n1} only cover the portion $X^2 -
T^2 > 0$, $X>0$ of flat space, which is called a Rindler wedge.

\begin{figure}[htbp]
\centering
\psfrag{T}{T}
\psfrag{X}{X}
\psfrag{L}{L}
\psfrag{R}{R}
\psfrag{xconst}{$x=$ constant}
\psfrag{tconst}{$t=$ constant}
\includegraphics[width=0.5\textwidth]{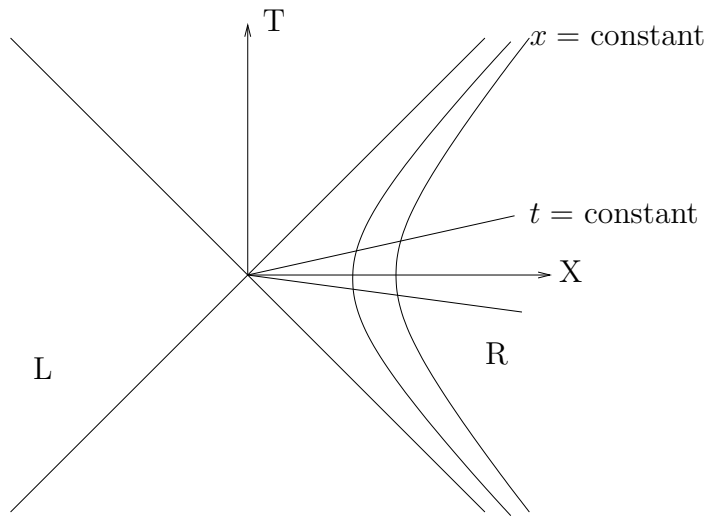}
\caption{Rindler horizon in flat space.}
\label{fig4}
\end{figure}

We see that there is a Killing horizon in flat space, as shown in
figure~\ref{fig4}. The surface at $x=0$ in the Rindler coordinates is
a Killing horizon for the Killing vector $\xi = \partial_t$. This
horizon is called a Rindler horizon. The Killing vector $\xi =
\partial_t$ in the Rindler coordinates is recognised as a boost
symmetry, and the co-moving observers in Rindler coordinates are
uniformly accelerating, with acceleration $a= 1/x$. The horizon is
therefore also referred to as an acceleration horizon. This horizon
arises because these uniformly accelerating observers will not be able
to receive signals from the whole spacetime. The Rindler observers,
who follow worldlines $x=$ constant, cannot influence or be influenced
by events in the left wedge in figure~\ref{fig4}. Of course, since
this is just flat space, this is not an event horizon; it is an
example of an observer-dependent horizon, specified by reference
to a particular observer's worldline.

The Rindler horizon is the simplest example of a Killing horizon, and
provides a template for any investigation of Killing horizons. We can
extend the above analysis to show that the near-horizon limit of an
arbitrary bifurcate Killing horizon is a Rindler horizon. The fact
that we can in this sense reduce the local properties of any bifurcate
Killing horizon to considering a Rindler horizon will play a central
role in the universality of the thermodynamic properties of black
holes.  

\section{Classical black hole thermodynamics}
\label{class}

In this lecture, we will discuss the features of classical black hole
mechanics which provided the first indications that there was a
relation between black holes and thermodynamics. We aim to give a
qualitative description, providing a flavour for the results and a few
illustrative examples. For details of proofs and fuller arguments, we
refer the reader to the references; the lecture notes on black holes
by Townsend~\cite{town} include proofs of many of the statements we
quote here.

\subsection{Area theorem}

We consider first Hawking's area theorem~\cite{harea} and the relation
to the second law of thermodynamics. The area theorem states: \textit{
If the spacetime on and outside the future event horizon is a regular
predictable space, and the stress tensor satisfies the null energy
condition, $T_{\mu\nu} k^\mu k^\nu \geq 0$ for arbitrary null $k^\mu$,
then the area of spatial cross-sections of the event horizon is
non-decreasing. } The condition that the spacetime is a regular
predictable space essentially forbids naked singularities on or
outside the event horizon; see~\cite{hawkellis} for details. Thus, so
long as spacetime is regular, and the matter satisfies an energy
condition, the area of the event horizon is non-decreasing. Bekenstein
pointed out~\cite{bek} that there was a close analogy between this
result and the second law of thermodynamics, and used it and
thermodynamic considerations to argue that black holes should be
assigned an entropy proportional to the area of the event horizon.

\begin{figure}[htbp]
\centering
\psfrag{th}{$\theta<0$}
\psfrag{p}{$p$}
\psfrag{q}{$q$}
\includegraphics{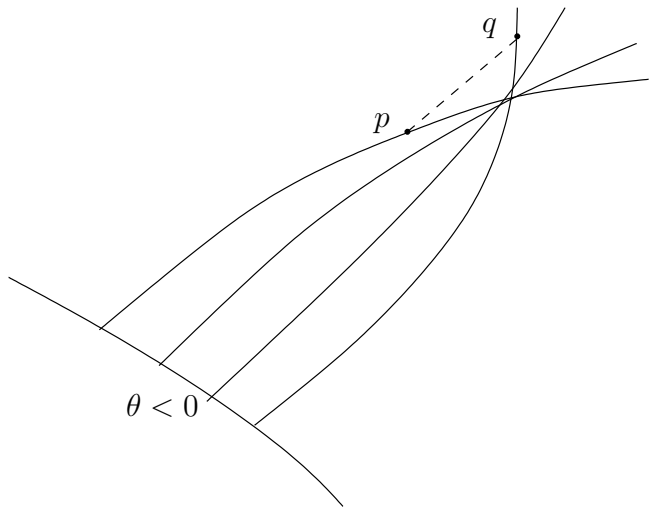}
\caption{A family of null geodesics with $\theta <0$ initially will
  form a caustic; the dotted curve connecting $p$ and $q$ lies within
  the local light cone, so these points are timelike separated.}
\label{fig5}
\end{figure}

Let us briefly give an idea of how this theorem is proved. If we
consider a small bundle of the null geodesics generating the event
horizon, which has a cross-sectional area $A$ at some value of the
affine parameter $\lambda$ along the geodesics, then we can define the
expansion $\theta$ by
\begin{equation}
\frac{dA}{d\lambda} = \theta A;
\end{equation}
that is, $\theta$ is the fractional rate of change of the area.  If we
imagine the theorem is violated, so that the area of the horizon
decreases, then we must have $\theta < 0$ somewhere on the event
horizon. Since the generators are geodesics, the evolution of the
expansion is determined by Raychaudhuri's equation,
\begin{equation}
\frac{d \theta}{d \lambda} = -\frac{1}{3} \theta^2 - \sigma_{\mu\nu}
\sigma^{\mu\nu} + \omega_{\mu\nu} \omega^{\mu\nu} - R_{\mu\nu} k^\mu k^\nu,
\end{equation}
where $\sigma_{\mu\nu}$ is the shear, $\omega_{\mu\nu}$ is the
rotation, which vanishes for the generators of a hypersurface, and
$k^\mu$ is the tangent to the null geodesics. Hence, if the null
energy condition is satisfied, so $R_{\mu\nu} k^\mu k^\nu \geq 0$,
$\theta < 0$ implies that $\theta \to -\infty$ in finite
$\lambda$. This produces a caustic, as shown in figure~\ref{fig5}. But
the points $p$ and $q$ indicated on the figure are timelike
separated. This contradicts our assumption that the null curves are
the generators of an event horizon, as no two points on the event
horizon can be timelike separated. Thus, by contradiction, the
cross-sectional area of an event horizon cannot decrease. It is
interesting to note that although the proof assumes Einstein's
equations, they are not used in an essential way, whereas they will be
in proving the first law. The assumption that the geometry is regular
on and outside the event horizon will continue to play an important
role throughout our discussion. 

\subsection{Zeroth and first laws}
\label{first}

The area theorem is the only one of the classical results which truly
concerns the dynamics of black hole event horizons. The zeroth and first
laws of black hole mechanics are concerned with equilibrium or
quasi-equilibrium processes. That is, they concern stationary black
holes, or adiabatic changes from one stationary black hole to
another. We will assume that for such black holes, the event horizon
is also a Killing horizon, using one of the arguments at the end of
section~\ref{hors}---this implies some limitations on the generality
of the following statements.

\begin{figure}[htbp]
\centering
\includegraphics[width=0.5\textwidth]{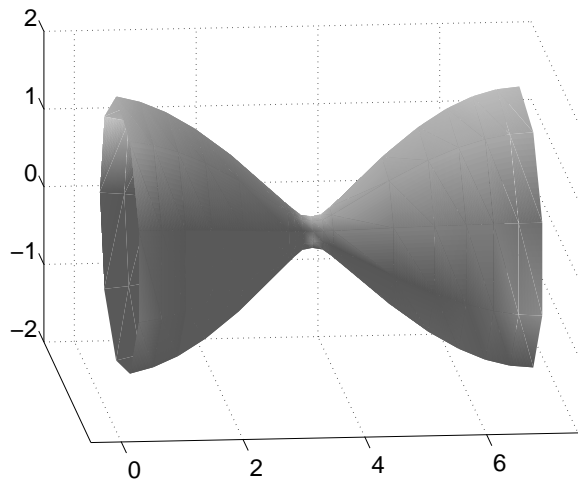}
\caption{The event horizon of a non-uniform black string solution,
  from~\cite{wiseman}.}
\label{wisefig}
\end{figure}

The zeroth law of black hole mechanics then states that the surface
gravity $\kappa$ is constant over the event horizon of a stationary
black hole~\cite{bch}. We have already seen that this is true for
Schwarzschild and Schwarzschild-AdS. This provides a first indication
that the surface gravity is an analogue of the temperature. This may
seem a weak analogy, since there are presumably many constant
quantities in a stationary black hole solution. Nonetheless, this is a
non-trivial statement. If we consider for example the non-uniform
black string solution which was discovered in~\cite{wiseman}, whose
horizon is depicted in figure~\ref{wisefig}, many local features of
the event horizon, such as its local radius of curvature, vary over
the horizon, but the surface gravity is constant by virtue of the
above result.

The analogy is considerably strengthened by the first law of black hole
mechanics. For a rotating charged black hole, this states that
\begin{equation} \label{firstl}
dM = \frac{\kappa}{8\pi} dA + \Omega dJ + \Phi dQ,
\end{equation}
where $\Omega$ is the angular velocity and $\Phi$ is the electric
potential~\cite{bch}.  This relates the change in area to the change
in mass, angular momentum and charges for an adiabatic transition
between nearby stationary black hole solutions. This first law is
evidently analogous to the first law of thermodynamics. Taken together
with the second law of thermodynamics, which identifies a multiple of
the area as the analogue of entropy, it enables us to identify a
multiple of the surface gravity $\kappa$ as an analogue of
temperature. Later, we reverse this argument: we can identify $T=
\frac{ \hbar \kappa}{2 \pi}$ as the temperature of a black hole using
quantum field theory in curved spacetime. The first law then implies
that $S = \frac{A}{4 G \hbar}$ {\it is} the black hole's entropy.

If we consider for example the Schwarzschild family of solutions, it
is straightforward to prove (\ref{firstl}) by direct calculation: the
area of the event horizon is $A = 16 \pi M^2$, so under a small change
$M \to M + dM$, $A \to 16 \pi (M + dM)^2 \approx A + 32 \pi M
dM$. Recalling that $\kappa = 1/4M$, we see that indeed $dM = \kappa
dA/ 8\pi$.

Similarly, if we consider the Schwarzschild-AdS black hole, for which 
\begin{equation}
A = \mbox{Vol}(S^{d-2}) r_+^{d-2},
\end{equation}
then
\begin{equation}
dA = \mbox{Vol}(S^{d-2}) (d-2) r_+^{d-3} dr_+,
\end{equation}
and we can relate $dr_+$ to $dM$ by observing that 
\begin{equation}
\begin{aligned}
0 &= f_{M+dM}(r_+ + dr_+) = f_{M+dM}(r_+) + dr_+ f'_{M}(r_+) \\
&= \left( 1 - \frac{\omega_d (M+dM)}{r_+^{d-3}} + \frac{r_+^2}{l^2}
\right) + dr_+ f'_{M}(r_+) \\
&= - \frac{\omega_d dM}{r_+^{d-3}} + dr_+ f'_{M}(r_+)
\end{aligned}
\end{equation}
Hence,
\begin{equation}
dM = \frac{r_+^{d-3} f'(r_+)}{\omega_d} dr_+ = \frac{f'(r_+)}{16 \pi}
(d-2)  \mbox{Vol}(S^{d-2}) r_+^{d-3} dr_+ = \frac{\kappa}{8\pi} dA,
\end{equation}
where we have used $\omega_d = 16 \pi/(d-2)\mbox{Vol}(S^{d-2})$ and
$\kappa = f'(r_+)/2$. Both these demonstrations have used the
equations of motion implicitly, as we have used details of the form of
the solutions. 

For more general asymptotically flat rotating charged black holes, it
is convenient to prove the first law by using the Komar integrals to
define $M$, $J$, and $Q$. As with the short proof above for the
Schwarzschild and Schwarzschild-AdS cases, the proof using Komar
integrals uses the equations of motion in an essential way.

Since $M$, $J$, $Q$ are defined asymptotically, the form of the first
law derived in this way is relating quantities defined at infinity to
quantities defined locally on the black hole horizon. This seems to
depend in a crucial way on the black hole uniqueness results, which
ensure that there is a unique spacetime geometry, and hence a
well-defined value of $A$, for given $M,J,Q$. In higher dimensions,
the solutions are no longer unique~\cite{er}, and can involve `dipole
charges' which are not globally conserved, and hence are not given by
asymptotic integral expressions~\cite{infnon}. It was shown
in~\cite{infnon} that these latter solutions satisfied a first law
including a work term associated with the dipole charge. Understanding
the first law for these cases in general is an interesting open
problem.

A potentially relevant development is the recent definition of a form
of the first law which is valid locally on the horizon, by introducing
a new notion of a horizon called an isolated
horizon~\cite{isol}. This also allows discussion of thermodynamics of black
holes assuming that only the black hole horizon is in equilibrium, not
necessarily the whole spacetime.

Another important generalisation, which has already played an
important role in string theory discussions, is the investigation by
Wald and Iyer~\cite{wald,iywald} (building on ideas from
\cite{sudwald,jacmy}), who showed that a first law holds for an
arbitrary Lagrangian theory of gravity, where the gravitational
Lagrangian $\mathbf{L}(R_{\mu\nu\rho\sigma}, \nabla_{\lambda}
R_{\mu\nu\rho\sigma}, \ldots)$ is some general functional of the
Riemann tensor and its covariant derivatives---this could be, for
example, the effective action for a string theory including $\alpha'$
corrections.

This approach uses a Noether current $\mathbf{J}$ associated with a
vector field $\xi$ constructed from a diffeomorphism variation of the
Lagrangian. When the equations of motion are satisfied, this current
defines an $(n-2)$ form charge $\mathbf{Q}[\xi]$ on spacetime
satisfying $ \mathbf{J} = d\mathbf{Q}$. They showed that if one
considered a stationary black hole solution in this theory, with a
bifurcate Killing horizon $\mathcal H$, one could relate an energy
$\mathcal{E}$ defined by a surface integral at infinity to the change
in a local integral over the horizon,
\begin{equation} \label{wald1}
\delta \mathcal{E} = \frac{\kappa}{2 \pi} \delta S + \mbox{ work terms},
\end{equation}
where $\kappa$ is the surface gravity defined as before, and the
quantity playing the role of entropy is now
\begin{equation} \label{genent}
S = 2 \pi \int_{\Sigma} \mathbf{Q}[\bar{\xi}] = 2\pi
\int_\Sigma \mathbf{X}^{cd} \epsilon_{cd}.
\end{equation}
In the expression for $S$, $\Sigma$ is the bifurcation surface in the
bifurcate Killing horizon $\mathcal H$, and $\bar{\xi}$ is the Killing
vector field which vanishes on this bifurcation surface. The second
form of the expression is obtained from the explicit form of the
Noether charge. The $n-2$ form $\mathbf{X}$ can be defined by taking a
functional derivative of the Lagrangian with respect to the Riemann
tensor,
\begin{equation}
\mathbf{X}_{ab} = - \frac{\delta \mathbf{L}}{\delta
  \mathbf{R}_{ab}},
\end{equation}
where $\mathbf{R}_{ab}$ is the usual curvature 2-form,
$R_{ab \mu\nu} = e_a^\lambda e_b^\rho R_{\lambda \rho
\mu\nu}$.  

This construction of the first law provides a notion of entropy for a
general gravitational theory. If the Lagrangian is simply the
Einstein-Hilbert Lagrangian, the form $\mathbf{X}_{ab}$ will be
proportional to the area element on the horizon, and this will reduce
to $S = A/4$. In general, the entropy is given by a local, geometrical
integral over the event horizon of the black hole, but this integral
may now involve the curvature as well as the proper area of the
horizon. Since~\eqref{wald1} is proved for a general perturbation in
the space of solutions, this formulation may also be sufficiently
general to encompass the first law for black rings with dipole
charges.

This formula for the entropy has been applied to the calculation of
the entropy for extremal black holes in string theory, including
higher-curvature
corrections~\cite{dewit1,dewit2,dewit3}.\footnote{These extremal black
  holes have $T=0$, and their horizons are not bifurcate Killing
  horizons. To apply the formula~\eqref{genent} to calculate their
  entropy, we must assume that the entropy is continuous as we move
  from the non-extreme black holes, where the argument of~\cite{wald}
  applies, to these extreme solutions.} Remarkably,
the changes in the entropy formula agree precisely with the results of
microscopic calculations from an M-theoretic
perspective~\cite{msw,vafa}. More recently, it has been suggested that
the partition function for these black holes, including all the
higher-curvature corrections, is related to the partition function for
topological strings~\cite{osv}.

An obvious next question is whether this entropy satisfies a second
law; that is, can one show that \eqref{genent} is always
non-decreasing in arbitrary physical processes, proving an analogue of
Hawking's area theorem for this more general expression? For
quasi-stationary processes, the second law can be shown to hold as a
consequence of the first law and the null energy condition \cite{jkm};
however, for general dynamical processes, the question remains open,
although a proof for a particular class of Lagrangians was given
in~\cite{jkm}. 

Already in this generalisation of the classical theory, we see that
the form of the temperature is unchanged, while the form of the
entropy appearing in this formula depends on the theory we
consider. The primary nature of the identification of the temperature
with surface gravity in \eqref{bhthermo} will become still more
apparent when we consider quantum effects in the next two sections.

In this section, we have seen that there is an analogy between the
classical laws of black hole mechanics and the laws of
thermodynamics. However, at the classical level, this is purely an
analogy---in particular, classical black holes are not thermal, as
they have no emission. To promote this analogy to a true
identification, we need to introduce a quantum element into our
discussion, by studying quantum field theory on the black hole
background. 

\section{Quantum field theory in curved spacetime}
\label{qft}

The subject of quantisation of the matter fields on a fixed spacetime
background is of clear practical importance. It provides the quantum
analogue of the classical kinematics of gravity determined by for
example the geodesic equation, and like the geodesic equation, it
should provide a good description in contexts where we are interested
in the effects of spacetime on some matter, but the gravitational
field of this matter itself is negligible. It is also analogous to
quantising the fields describing charged particles in a fixed
classical background electromagnetic field.

Unlike quantisation in a fixed background gauge field, quantisation on
curved spacetime involves significant conceptual differences from the
more familiar case of quantisation in flat space. This is because in
the usual treatment of quantum field theory, Poincare-invariance plays
a central role in the discussion. In considering a general spacetime
background, we have to give up on this approach based on global
symmetry. This means in particular that there will not be a preferred
vacuum state or a natural Fock space construction of the Hilbert
space. We are led to make a strong distinction between the
quantisation of the field, promoting it to a local operator, and
issues to do with the choice of state. 

In this section, I will briefly introduce the essential aspects of
this quantisation. The initial review of canonical quantisation will
follow the treatment in~\cite{wald94,jac} quite closely, adopting a
modern point of view where the focus is on promoting fields to local
operators and the invariant characterisation of states, rather than
mode decompositions and Fock spaces. I will also briefly discuss the
Peierls bracket, which provides an alternative, more covariant
approach to quantisation. Since our focus is on thermodynamics, I will
spend some time explaining the KMS condition, which provides a
characterisation of thermal states for general quantum systems, and
its application to a quantum field on a stationary background
spacetime.

Our discussion will be restricted to free theory, and it will be
sufficient for our purposes to consider a real scalar field. The
extension to other spins is fairly straightforward. So we start from a
classical theory with action
\begin{equation}
S = - \frac{1}{2} \int d^d x \sqrt{-g} (g^{\mu\nu} \nabla_\mu \varphi
\nabla_\nu \varphi + m^2 \varphi^2),
\end{equation}
with corresponding equation of motion
\begin{equation}
(\Box - m^2) \varphi =0.
\end{equation}

\subsection{Canonical quantisation}

If we assume our spacetime manifold $M$ can be decomposed as $M =
\mathbb{R} \times \Sigma$ where $\mathbb{R}$ represents the timelike
direction and $\Sigma$ is some $d-1$-dimensional Riemannian manifold,
then we can apply the usual techniques of canonical
quantisation. Denoting the corresponding decomposition of a coordinate
chart on $M$ by $x^\mu = (x^0, \vec{x})$, we can write
\begin{equation}
S = \int dx^0 L,
\end{equation}
and introduce as canonical coordinates the value of $\varphi$ on
$\Sigma$ at some value of $x^0$, $\varphi(\vec{x})$, and the conjugate momentum
\begin{equation}
\pi(\vec{x}) = \frac{\delta L}{\delta (\partial_0 \varphi)} = \sqrt{-g}
g^{\mu0} \partial_\mu \varphi = - \sqrt{h} n^\mu \partial_\mu
\varphi(\vec{x}),
\end{equation}
where $h_{ij}$ is the metric on $\Sigma$ and $n_\mu$ is the normal to
$\Sigma$ in $M$. 

We then quantise by imposing the usual canonical
commutation relations,
\begin{equation}
[\varphi(\vec{x}), \pi(\vec{x})] = i \hbar \delta^{d-1}(\vec{x},\vec{y}),
\end{equation}
where the densitized delta function $\delta^{d-1}(\vec{x},\vec{y})$ is
defined by
\begin{equation}
\int_\Sigma d^{d-1} \vec{x} f(\vec{y})\delta^{d-1}(\vec{x},\vec{y}) =
f(\vec{x}). 
\end{equation}
This is the essence of quantisation. It is worth noting that this step
is entirely local. The overall structure of the spacetime does not
enter into the definition of these commutation relations. It is only
when we wish to go on to represent them in terms of operators acting
on a Hilbert space that the structure of the background becomes
important, entering notably through the definition of the inner
product. 

The usual flat-space definition of this Klein-Gordon inner product
generalises directly to curved spacetime,
\begin{equation} \label{innp}
(f, g)=  \int_\Sigma d\Sigma_\mu j^\mu(f,g),
\end{equation}
where the current
\begin{equation}
j^\mu(f,g) = -i \sqrt{-g} g^{\mu\nu} (\bar{f} \partial_\nu g -
\partial_\nu \bar{f} g). 
\end{equation}
This current is conserved as a consequence of the equations of motion,
$\partial_\mu j^\mu= 0$, which implies that the inner product $(f, g)$
is independent of the choice of spacelike slice $\Sigma$; that is, it
is in particular independent of the time coordinate $x^0$.  As a
consequence of the definition, this inner product also satisfies
\begin{equation}
\overline{(f,g)} = - (\bar{f},\bar{g}) = (g,f), 
\end{equation}
and as a result, $(f,\bar{f}) =0$. The inner product is clearly not
positive definite.

So far, the quantisation procedure has precisely parallelled the usual
discussion. At this stage in flat space, we would introduce a basis of
positive frequency field modes. The restriction of the inner product
\eqref{innp} to these modes would then be positive definite, and we
could use it to construct the Hilbert space. The essential difference
in a curved spacetime is that there is no natural \textit{a priori}
notion of positive frequency. To construct a Hilbert space, we
therefore need to introduce a decomposition of the space $\mathcal{S}$ of
solutions of the field equations into a positive norm part and its
conjugate:
\begin{equation}
\mathcal{ S} = \mathcal{ S}_p \oplus \bar{\mathcal{ S}_p},
\end{equation}
where
\begin{equation}
(f,f) > 0 \quad \forall f \in \mathcal{ S}_p, \qquad (f, \bar{g}) = 0
  \quad \forall f,g
  \in \mathcal{ S}_p. 
\end{equation}
Given such a choice of decomposition, we can define annihilation and
creation operators for the mode $f$ by
\begin{equation}
a(f) = (f, \varphi), \quad a^\dagger(f) = -a(\bar{f}) = -
(\bar{f},\varphi). 
\end{equation}
The canonical commutation relations imply that these operators will
satisfy the usual algebra of creation and annihilation operators,
\begin{equation}
[a(f), a^\dagger(g)] = (f,g), \quad [a(f),a(g)] =
[a^\dagger(f),a^\dagger(g)] = 0.
\end{equation}
We may, if we wish, identify a vacuum state associated with this
decomposition of the space of solutions by
\begin{equation}
a(f) |0 \rangle = 0 \quad \forall f \in \mathcal{ S}_p,
\end{equation}
and construct a Fock space by taking the span of all states of the
form
\begin{equation}
a^\dagger(f_{i_1}) \ldots a^\dagger(f_{i_n}) | 0 \rangle
\end{equation}
for all $n$ and all $f_{i_k} \in \mathcal{ S}_p$.

Given a choice of positive norm subspace $\mathcal{ S}_p$, we can
  construct an orthonormal basis $f_n \in \mathcal{ S}_p$. Writing the
  annihilation and creation operators associated to this basis as $a_n
  = (f_n,\varphi)$, $a^\dagger_n = -(\bar{f}_n, \varphi)$, we can then
  make contact with the standard treatment of quantisation by
  introducing the usual mode decomposition of the field operator
\begin{equation}
\varphi = \sum_n (a_n f_n + a^\dagger_n \bar{f}_n), 
\end{equation}
and the usual $n$-particle basis of states for the Fock space,
\begin{equation}
a^\dagger_{n_1} \ldots a^\dagger_{n_k} |0\rangle. 
\end{equation}

The essential difficulty in this construction is the specification of
a positive norm subspace $\mathcal{ S}_p \subset \mathcal{ S}$, which
corresponds physically to the specification of a notion of positive
frequency. It is always possible to find such subspaces, but in a
general curved spacetime the background structure of the spacetime
does not select any particular one as a natural candidate for defining
the vacuum and Fock space. It can be shown~\cite{wald94} that the
Fock space constructions based on different notions of positive
frequency will be unitarily inequivalent. That is, different
constructions are giving genuinely different representations of the
algebra of field operators, and the vacuum constructed with respect to
one notion of positive frequency  $\mathcal{ S}_p$, $|0\rangle_p$,
will not lie in the Fock space built on the vacuum $|0\rangle_{p'}$
associated with a different choice $\mathcal{ S}_{p'}$. 

There is a linear relation between different choices: any
$f' \in \mathcal{ S}_{p'}$ can be written as $f' = f + \bar{g}$ for
some $f, g \in \mathcal{ S}_p$. If we introduce bases, this linear
relation can be expressed through the Bogoliubov transformation
\begin{equation}
f'_n = \sum_{m} \alpha_{nm} f_m + \beta_{nm} \bar{f}_m.
\end{equation}
The mixing of positive and negative frequency is expressed through the
presence of the coefficients $\beta_{nm}$, which could not appear in
the transformation between different bases for the same $\mathcal{
  S}_p$. 

\subsection{Peierls bracket}

The use of a canonical decomposition of spacetime, splitting it into a
Riemannian spatial manifold $\Sigma$ and a time direction, appears
contrary to the spirit of special and general relativity. As Minkowski
presciently observed, ``space by itself, and time by itself, are
doomed to fade away into mere shadows and only a kind of union of the
two will retain an independent reality''~\cite{minkquot}. While it has
been shown that the canonical quantisation described above is
independent of the particular foliation of spacetime we choose, it
would be preferable to have a covariant method which avoided the need
to introduce a foliation altogether.

Fortunately, such a method exists. It is due to
Peierls~\cite{peierls}. The key idea of the method is to introduce a
bracket between classical observables which is constructed entirely
covariantly, but which is equivalent to the Poisson bracket of the
conventional canonical treatment.  The Peierls bracket formalism and
covariant quantisation is described in great detail and generality
in~\cite{dewitt03}. There is also a useful brief discussion
in~\cite{haag}.

Consider classical observables $A,B$, which can be integrals of some
arbitrary local functions of the fields and derivatives over a region
of spacetime. To construct their Peierls bracket, consider adding an
infinitesimal contribution to the Lagrangian
\begin{equation}
\mathcal{L} \to \mathcal{L} + \epsilon A.
\end{equation}
This will produce a corresponding change in the solutions of the
equations of motion: we consider the change with advanced or retarded
boundary conditions $ \delta^\pm_A \varphi$, which vanish respectively
in the region to the future or past of the perturbation $A$. These
define corresponding changes $\delta^\pm_A B$ in the other observable
$B$.

The difference between the advanced and retarded solutions,  
\begin{equation}
\delta_A \varphi = \delta^-_A \varphi - \delta^+_A \varphi, 
\end{equation}
defines an action of $A$ on the space of classical solutions, as
$\varphi + \epsilon \delta_A \varphi$ satisfies the equations of
motion of the original Lagrangian $\mathcal{L}$ to first order in
$\epsilon$. The Peierls bracket is then defined as the action of $A$
on $B$:
\begin{equation} \label{pbrac}
(A, B) \equiv \delta_A B = \delta^-_A B - \delta^+_A B. 
\end{equation}
There is also a natural reciprocity relation, $\delta_A B = -\delta_B
A$, which implies that this can also be expressed as the action of $B$
on $A$. Note that because $\delta_A \varphi$ defined a motion in the
space of classical solutions, the Peierls bracket is only defined
on-shell; that is, \eqref{pbrac} only defines a bracket between the
observables $A$, $B$ evaluated on some classical solution. 

In our example of a free real scalar field, the Peierls bracket of the
fundamental fields is
\begin{equation}
(\varphi(x), \varphi(x')) = i \Delta(x-x'),
\end{equation}
where $\Delta(x-x')$ is the difference between the retarded and
advanced propagators. This is the same as the Poisson bracket of the
fields, establishing for this particular case the general equivalence
of the Peierls bracket and the Poisson bracket. We can quantise the
theory covariantly by simply promoting the fields to operators and
imposing the Peierls bracket as a commutator relation on the field
operators:
\begin{equation}
[ \varphi(x), \varphi(x')] = i \Delta(x-x').
\end{equation}
This produces the same algebra of observables as the canonical
quantisation while avoiding the need to introduce any canonical
structures.

\subsection{Observables \& Hadamard states}

We distinguish different states in quantum field theory by the
expectation values they give for certain operators. In flat spacetime,
states are labelled with the value of the particle number operator $N
= \sum_n a_n^\dagger a_n$. Intuitively, states with larger $\langle N
\rangle$ are `farther' from the vacuum. In a general curved spacetime,
however, where there is no natural notion of positive frequency, this
value depends on an arbitrary choice we are making when we define
$a_n$, so it cannot be a physically meaningful quantity. There is no
invariant notion of the number of particles in the space, even at a
fixed moment in time.

We therefore want to label states not by particle number, but with the
values for physical observables. A class of simple operators we can
consider are the time-ordered products of the fundamental field
operator. The expectation values of these operators are sufficient to
distinguish between different states. That is, we consider the
$n$-point expectation values of fundamental field operators, $\langle
\varphi(x) \varphi(y) \rangle$, $\langle \varphi(x) \varphi(y)
\varphi(z) \rangle$, etc. In our particular case of a free scalar
field, a state can be specified by specifying the two-point
function,\footnote{The states where the one-point function vanishes
and higher-point functions are determined by the two-point function
are termed quasi-free~\cite{wald94}. We will assume henceforth that
the state is quasi-free, so that the two-point function suffices to
uniquely specify the state.}
\begin{equation}
G(x,y)= \langle \varphi(x) \varphi(y) \rangle.
\end{equation}

However, there are quantities we are physically interested in which
are not the expectation value of any operator of this form. For
gravitational purposes, the most important example is the
stress-energy tensor $\langle T_{\mu\nu} \rangle$, the source of the
gravitational field in a semi-classical treatment. The classical
stress-energy tensor cannot be simply promoted to a quantum operator,
because it involves the product of fields at the same spacetime point,
and these products are not well-defined for the distributional field
operators of the quantum theory. 

In flat spacetime, we deal with this problem by normal ordering,
rewriting the stress tensor in terms of the annihilation and creation
operators $a_n, a_n^\dagger$ (whose product is well-defined), with a
particular choice of operator ordering, which corresponds to measuring
the difference between $\langle T_{\mu\nu} \rangle$ in the state of
interest and its value in the vacuum. However, once again, such an
expression in terms of creation and annihilation operators will not
produce invariant results in curved spacetimes.

The question is then, is there some alternative approach to the
construction of $\langle T_{\mu\nu} \rangle$?  Remarkably, it has been
shown (see~\cite{wald94} for details) that one can construct a local,
conserved stress-energy tensor $\langle T_{\mu\nu} \rangle$, but only
for a certain class of states, called Hadamard states.

Hadamard states are defined by the requirement that at short
distances, the two-point function behave as
\begin{equation}
\lim_{x \to y} G(x,y) = \frac{U(x,y)}{4\pi^2 \sigma(x,y)} + V(x,y) \ln
\sigma + W(x,y),
\end{equation}
where $\sigma$ is half the square of the geodesic distance between $x$
and $y$, and $U$, $V$, and $W$ are smooth functions with $U(x,x)=
1$. The form of the functions $U(x,y)$ and $V(x,y)$ is then
determined by requiring that the two-point function satisfies the
Klein-Gordon equation in $x$.
This restriction on the form of the two-point function can be
motivated by observing that it is satisfied by the two-point function
in any state in the Fock space built on the usual Minkowski vacuum in
flat space. Thus, the restriction to Hadamard states corresponds to
the reasonable assumption that when we consider the behaviour on
scales much smaller than any curvature scale, using our usual
flat-space techniques will be a good approximation.

Thus, the short-distance singularity in Hadamard states is entirely
determined by the local spacetime geometry. They are the closest we
can come in a curved spacetime to our usual Fock space states built on
the Minkowski vacuum: the constraint on the short distance
singularities will mean that all physical observers will see a finite
density of particles in any Hadamard state.
We can remove this ultraviolet divergence by a local and
state-independent renormalisation procedure. The renormalised
two-point function can be used to construct $\langle T_{\mu\nu}
\rangle$, giving a unique result up to the addition of local curvature
terms (which represent a renormalisation of the gravitational
couplings)~\cite{wald94}.

The Hadamard condition is preserved by evolution, so if the state is
initially Hadamard in a suitable neighbourhood of some Cauchy surface,
it will be Hadamard throughout the spacetime~\cite{kaywald}. It has
also been shown that if the spacetime is spatially compact, any
Hadamard state can be constructed as a state in the Fock space defined
on any other Hadamard state~\cite{wald94}. These considerations all
motivate adopting the Hadamard condition as a natural condition on
states.

\subsection{KMS condition}

In this subsection, we will finally describe the general
characterisation of thermal states in quantum field theory; this is
provided by the KMS condition, named for Kubo~\cite{kubo} and Martin
and Schwinger~\cite{ms}. We first give a brief description of the KMS
condition, motivating it by relating it to the usual notion of thermal
equilibrium for a system with finitely many degrees of freedom. We
then discuss the application of this condition to a free scalar field,
and show that thermal states of a free scalar field are characterised
by having a two-point function which is `periodic in imaginary
time'. More detailed discussions of the KMS condition can be found in
\cite{haag,sewellbook}. My discussion of the KMS condition for a
scalar field follows~\cite{fullr}.

In a system with finitely many degrees of freedom, the standard
definition of the canonical ensemble describing thermal equilibrium at
temperature $T = \beta^{-1}$ is that for all operators $A$, the
expectation value is given by 
\begin{equation} \label{thermo}
\langle A \rangle_\beta = \frac{1}{Z} \mathrm{Tr}(e^{-\beta H} A )
\end{equation}
where $H$ is the Hamiltonian of the system and $Z =
\mathrm{Tr}(e^{-\beta H})$ is the canonical partition
function. However, for a system with a continuous spectrum, such as a
field theory in infinite volume, this can only provide a formal
definition of thermal behaviour, since this partition function
diverges (as will $\mathrm{Tr}(e^{-\beta H} A )$ for most operators of
interest).

A more general characterisation of thermality can be obtained by
considering the consequences of the above definition. Consider the
expectation value  
\begin{equation}
\langle A_t B\rangle = \langle e^{itH} A e^{-itH} B \rangle, 
\end{equation}
and regarding this as a function of time, extend the index $t$ to
imaginary values. Then, in the case of a simple quantum mechanical
system, we can deduce from \eqref{thermo} that
\begin{equation}
\langle A_{-i\beta} B \rangle_\beta = \frac{1}{Z} \mathrm{Tr}[e^{-\beta H}
  (e^{\beta H} A e^{-\beta H}) B] = \frac{1}{Z} \mathrm{Tr}[e^{-\beta H}
  B A] = \langle B A\rangle_\beta 
\end{equation}
for all bounded operators $A,B$, where we have used the cyclic
property of the trace in the second step. This provides us with a
condition that only makes reference to finite quantities,
the KMS condition~\cite{kubo,ms}
\begin{equation} \label{kms}
\langle A_{-i\beta} B \rangle_\beta = \langle B A\rangle_\beta. 
\end{equation}
Any state such that \eqref{kms} holds for all bounded operators $A,B$
is called a KMS state.

This condition has been derived as a consequence of \eqref{thermo},
but it provides a suitable notion of thermal states for more general
systems. It has been shown to be equivalent to \eqref{thermo} where
that definition applies. Furthermore, if an infinite system satisfies
the KMS condition, any finite system coupled to it will approach
thermal equilibrium in the sense of \eqref{thermo}; thus, a system
satisfying \eqref{kms} behaves as a thermal reservoir. Finally, a
system satisfying the KMS condition minimises the free energy
locally. For proofs of these statements, see \cite{haag,sewellbook}.

We will now specialise to the case of a free scalar field in curved
spacetime, and investigate the application of the KMS condition,
following~\cite{fullr} quite closely. We observe first that an
equilibrium state is by definition a stationary state, so for any
notion of equilibrium state to exist, the spacetime must admit a
timelike Killing vector field, with associated observers whose
worldlines are orbits of the Killing vector field; the notion of
equilibrium will be with respect to measurements by these
observers. Call this Killing vector $\partial_t$.

The observable of interest in the theory of a free scalar field is the
two-point function $\langle \varphi(x) \varphi(y) \rangle$; the
Killing symmetry implies that this only depends on the time
difference, $t_x-t_y$. We introduce the positive and negative
frequency Wightman functions
\begin{equation} \label{gplus}
G_+ (t_x - t_y, \vec{x}, \vec{y}) = \langle \varphi(x) \varphi(y)
\rangle 
\end{equation}
and
\begin{equation} \label{gminus}
G_- (t_x - t_y, \vec{x}, \vec{y}) = \langle \varphi(y) \varphi(x)
\rangle = G_+(t_x - t_y, \vec{x}, \vec{y}) + [\varphi(x),\varphi(y)],
\end{equation}
where the $c$-number commutator is the same for all states. 

\begin{figure}[htbp]
\centering
\psfrag{t}{$t$}
\psfrag{b}{$\beta$}
\includegraphics[width=0.7\textwidth]{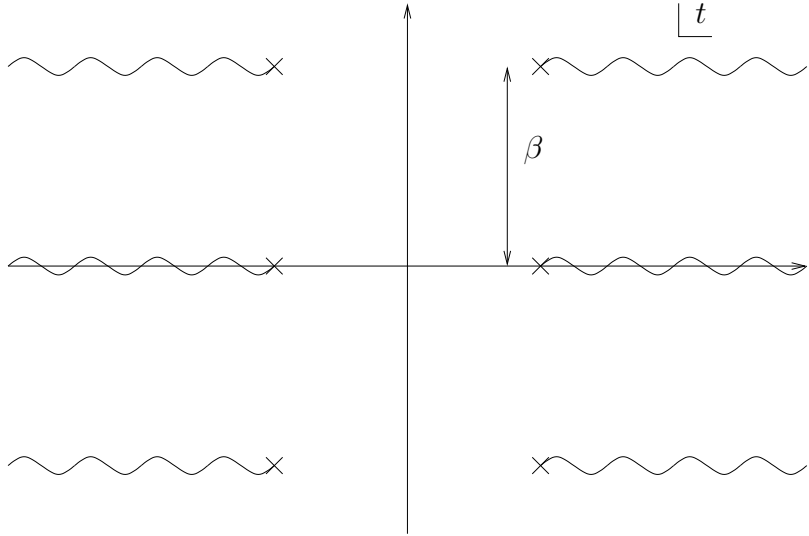}
\caption{The pole structure of $\mathcal{G}^\beta (t, \vec{x},
  \vec{y})$ for a massive scalar field in the complex $t$ plane for
  fixed values of $\vec{x}\neq \vec{y}$. The crosses represent poles
  and the wavy lines are branch cuts.}
\label{fig6}
\end{figure}

Extending the dependence on $t_x-t_y$ into the complex plane, one can
show that $G_+^\beta (z, \vec{x}, \vec{y})$ is holomorphic below the
real axis, for $-\beta < \Im(z) <0$, while $G_-^\beta (z, \vec{x},
\vec{y})$ is holomorphic for $0 < \Im(z) < \beta$, where the
superscript implies that we take the expectation values on the RHS of
(\ref{gplus},\ref{gminus}) in a thermal state at temperature $T =
\beta^{-1}$. The KMS condition relates the two Wightman functions,
\begin{equation}
G_+^\beta (z - i\beta , \vec{x}, \vec{y}) = G_-^\beta (z, \vec{x}, \vec{y}). 
\end{equation}
There is another relation between them, which comes from the vanishing
of the commutator at spacelike separations: if $\vec{x} \neq \vec{y}$,
it follows from the second equality in \eqref{gminus} that at
sufficiently small $t= t_x - t_y$, $G_+ (t, \vec{x}, \vec{y}) = G_-
(t, \vec{x}, \vec{y})$. Using these two relations, $G_\pm^\beta(t,
\vec{x}, \vec{y})$ can be extended to define a single function on the
complex plane (for fixed $\vec{x}, \vec{y}$), $\mathcal{G}^\beta (z,
\vec{x}, \vec{y})$, which satisfies the periodicity
\begin{equation} \label{therper}
\mathcal{G}^\beta (z, \vec{x}, \vec{y}) = \mathcal{G}^\beta(z +
in\beta, \vec{x}, \vec{y})
\end{equation}
for $n \in \mathbb{Z}$, and is holomorphic in the complex $z$ plane
 away from the lines $\Re(z)^2 > |\vec{x}- \vec{y}|^2$, $\Im(z) = n
 \beta$. A picture of this pole structure for a massive scalar field
 is given in figure~\ref{fig6}. It is this characterisation of the
 thermal state in terms of the two-point function which is most useful
 and most commonly used in discussing quantum field theory in curved
 spacetime.

A thermal state for a free scalar field on a stationary spacetime
$\mathcal M = \mathbb{R}_t \times \Sigma$ is defined by a two-point
function $\mathcal{G}^\beta (z, \vec{x}, \vec{y})$ satisfying the
periodicity condition \eqref{therper}. In particular, if the spacetime
is static, so that the analytic continuation $t \to i \theta$ defines
a Euclidean spacetime $\mathcal M_E = \mathbb{R}_\theta \times
\Sigma$, Green's functions $G^\beta (\theta, \vec{x}, \vec{y})$ on the
Euclidean spacetime periodic under $\theta \to \theta + \beta$ are
related by analytic continuation to thermal states on the original
Lorentzian spacetime.

\section{Unruh \& Hawking radiation}
\label{hrad}

We now turn to the central task, demonstrating that event horizons
have thermal properties for a quantum field in curved spacetime. 

\subsection{Unruh radiation}

We consider first the thermal Unruh radiation associated to the
Rindler horizon in flat spacetime. This is the simplest example of a
bifurcate Killing horizon, and as discussed in section~\ref{rindler},
it provides a good approximation to the geometry in the neighbourhood
of a generator of an arbitrary Killing horizon. Hence, this case
contains the essential physics. In flat spacetime, we have a preferred
state for a quantum field, namely the Minkowski vacuum. We wish to
show that this state will appear thermal with respect to the Rindler
time. This result was originally derived by Unruh~\cite{unruh},
following Hawking's derivation of Hawking radiation
in~\cite{hrad1,hrad2}, and is therefore called the Unruh effect.

Recall that the Rindler coordinates $(x,t)$ were related to the usual
Cartesian coordinates $(T,X)$ on flat space by \eqref{rflat}, 
\begin{equation}
X = x \cosh \kappa t, \quad T = x \sinh \kappa t, 
\end{equation}
so we have
\begin{equation}
ds^2 = -dT^2 + dX^2 + ds^2_{\mathbb{R}^{(d-2)}} = - \kappa^2 x^2 dt^2
+ dx^2 +  ds^2_{\mathbb{R}^{(d-2)}}. 
\end{equation}
Another useful coordinate system can be defined by setting $x =
e^{\kappa \rho}$, so 
\begin{equation} \label{confr}
ds^2 = \kappa^2 e^{2\kappa \rho} (-dt^2 + d\rho^2) +
ds^2_{\mathbb{R}^{(d-2)}} = -2 \kappa^2 e^{2\kappa (u-v)} du dv +
ds^2_{\mathbb{R}^{(d-2)}},
\end{equation}
where $u,v = t \pm \rho$. These null coordinates are related to the
Cartesian ones $U,V = T \pm X$ by $U = e^{\kappa u}, V = - e^{-\kappa
  v}$. 

We consider the two-point function in the Minkowski vacuum, 
\begin{equation}
G_M(T_X-T_Y,X,Y) = \langle0 | \varphi(x^\mu) \varphi(y^\mu) | 0 \rangle.
\end{equation}
We could use the explicit form of this two-point function and apply
the above coordinate transformations directly to show that, written in
terms of the Rindler coordinates,
\begin{equation}
G_M(T_X-T_Y,X,Y) = G^{2\pi/\kappa}_R(t_x-t_y,x,y);
\end{equation}
that is, that the resulting function has period $2\pi/\kappa$ in the
complex $t_x - t_y$ plane. However, to avoid writing lengthy formulae
and to demonstrate the simplicity and inevitability of the result, we
will instead argue for this result more indirectly.

\begin{figure}[htbp]
\centering
\psfrag{t}{$t$}
\includegraphics[width=0.7\textwidth]{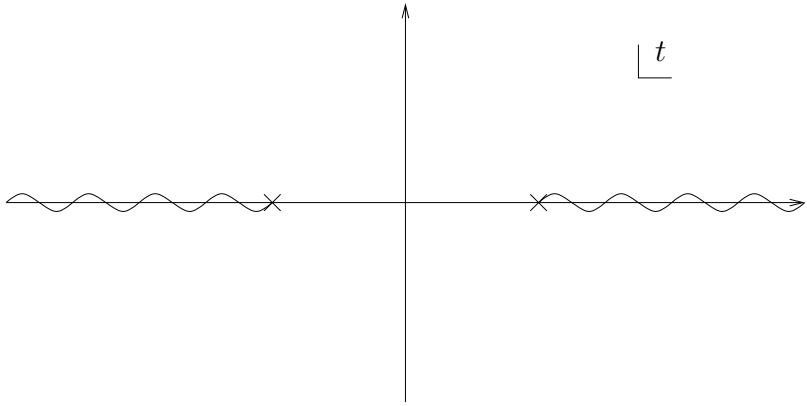}
\caption{The pole structure of $G_M$.  } 
\label{fig7}
\end{figure}

Consider the extension of $G_M(T_X-T_Y,X,Y)$ to a complex function in
the complex $Z = T_X - T_Y$ plane. This function will be holomorphic
for $\Im(Z) \neq 0$, as shown in figure~\ref{fig7}. In particular, if
we consider $Z = i(W_X-W_Y)$, this will define a regular Green's
function
\begin{equation}
G_E(W_X-W_Y,X,Y) = G_M(Z,X,Y) 
\end{equation}
satisfying
\begin{equation} \label{egreen}
(\Box_E - m^2) G_E = \delta^d(x^\mu-y^\mu)
\end{equation}
on the Euclidean flat space defined by analytically continuing $T \to iW$, 
\begin{equation}
ds^2 = dW^2 + dX^2 + ds^2_{\mathbb{R}^{(d-2)}}. 
\end{equation}
This is just the inversion of familiar statement that the vacuum
two-point function can be defined by Wick rotation from the Euclidean
Green's function. Depending on how we choose to approach the axis,
this defines the Feynman propagator, one of the Wightman functions, or
some other propagator, but in any case, the Euclidean Green's function
specified by \eqref{egreen} is uniquely defined.

Now we define new coordinates on the Euclidean space by $W = x \sin
\kappa \tau$, $X = x \cos \kappa \tau$, so 
\begin{equation}
ds^2 = \kappa^2 x^2 d\tau^2 + dx^2 + ds^2_{\mathbb{R}^{(d-2)}}. 
\end{equation}
Clearly $G_E(\tau_x-\tau_y,x,y)$ is a periodic function of $\tau_x
-\tau_y$ with period $2\pi/\kappa$. But this spacetime is also the
result of analytically continuing the Rindler time coordinate $t \to i
\tau$; that is, once we analytically continue to the Euclidean
section, the coordinate transformation \eqref{rflat} relating Rindler
and flat coordinates becomes the relation of Cartesian to polar
coordinates. The analytic continuation of the boost isometry which
defined the Rindler time translation is a rotation. Hence, if we now
extend $G_E(\tau_x-\tau_y,x,y)$ to a holomorphic function of a complex
coordinate $z = \tau_x -\tau_y$, we get
\begin{equation}
G_E(z,x,y)|_{z = -i(t_x-t_y)} = G^{2\pi/\kappa}_R (t_x-t_y,x,y),
\end{equation}
and we have verified that the Minkowski vacuum is indeed a thermal
state with respect to Rindler time with temperature $T =
\kappa/2\pi$. We see that from this point of view, the thermal
behaviour arises from the analytic continuation relation between
Rindler coordinates in Minkowski space and polar coordinates in the
Euclidean section. 

We have discussed the case of a free scalar field, but the result is
far more general: Bisognano and Wichmann~\cite{bw1,bw2} proved a
theorem which implies (as shown in~\cite{sewell}) that the Minkowski
vacuum satisfies the KMS condition with respect to the boost
time-translation Killing vector for an arbitrary interacting field
theory on flat space.

There is still an unanswered question, however: the Minkowski vacuum
is a pure state, so how can it be transmuted into a thermal state by
simply making a change of coordinates? This thermal behaviour actually
comes from the fact that a Rindler observer, following an orbit of
$\partial_t$, cannot see the whole spacetime. Thus, to describe the
state as seen by such an observer, we should trace over the degrees of
freedom in the unseen region of spacetime. This produces a mixed state
describing the Rindler observer's measurements, with entropy coming
from the entanglement between modes in the two sides of the spacetime.

To describe this entanglement, we need to relate the basis in the
space of solutions which is appropriate for a Rindler observer to the
usual Minkowski modes. The Minkowski modes are the usual plane-wave
modes
\begin{equation}
u_k = \psi_k e^{-i\omega_k T},
\end{equation}
where $\psi_k$ is a spatial wavefunction whose explicit form we do not
need, and $\omega_k$ is a positive frequency. The Minkowski vacuum
$|0\rangle$ is defined by $a_k | 0 \rangle =0$ for all $k$, where $a_k =
(u_k,\varphi)$ is the annihilation operator associated with the mode
$u_k$. The Rindler modes are defined by
\begin{equation}
{}^R u_k = \left\{ \begin{array}{l} \tilde \psi_k e^{-i\omega t}
  \quad \mbox{ in R,} \\ 0 \quad \mbox { in L,} \end{array} \right.
\end{equation}
\begin{equation}
{}^L u_k = \left\{ \begin{array}{l}  0
  \quad \mbox{ in R,} \\ \tilde \psi_k e^{i\omega t} \quad \mbox { in
  L,} \end{array} \right. 
\end{equation}
where L and R refer to the two wedges in figure~\ref{fig4}. 

The Minkowski vacuum contains correlations between the L and R Rindler
modes. Understanding the relation between the modes is simplified by
using an observation due to Unruh~\cite{unruh}, which emphasises the
physics of the correlations. For simplicity, I will describe this in
the case $m^2=0$; the extension to massive fields is left as an
exercise.

For $m^2=0$ the positive frequency Rindler mode in the R wedge is
simply\footnote{The mode has the same form as a plane wave mode in Cartesian
coordinates, as the R wedge is conformal to a flat space (as we can
see from~\eqref{confr}). The frequency with respect to the Rindler time
coordinate $t$ is $\omega$. However, unlike the usual plane wave
modes, the physical energy of the mode redshifts with $x$. That is,
the energy measured by a given Rindler observer will be $E_{local} =
\omega/(\kappa x)$. Hence, if we considered an outgoing wavepacket, which will
follow the null lines of constant $U$, it will redshift as it
propagates to larger $x$.}
\begin{equation}
p = \left\{ \begin{array}{l} e^{-i\omega v}
  \quad \mbox{ in R,} \\ 0 \quad \mbox { in L.} \end{array} \right.
\end{equation}
The mode $p$ is not purely positive frequency with respect to the
Minkowski time $T$. However, Unruh observed that we can construct a
solution which {\it is} purely positive frequency with respect to $T$,
and which agrees with $p$ in the R wedge. Write $p$ in Minkowski
coordinates, as
\begin{equation}
p = \left\{ \begin{array}{l} e^{i\omega \ln(-V)/\kappa}
  \quad \mbox{ for } V<0 \\ 0 \quad \mbox { for } V>0. \end{array} \right.
\end{equation}
Then 
\begin{equation}
P = \left\{ \begin{array}{l} e^{i\omega \ln(-V)/\kappa}
  \quad \mbox{ for } V<0 \\   e^{-\pi \omega/\kappa} e^{i\omega \ln(V)/\kappa} \quad
  \mbox { for } V>0 \end{array} \right. 
\end{equation}
is the required mode solution; it clearly agrees with $p$ for $V<0$,
and it is purely positive frequency, as it is analytic in the
lower-half $V$ plane. We can also rewrite this as
\begin{equation}
P = p + e^{-\pi \omega/\kappa} \tilde p,
\end{equation}
where 
\begin{equation}
\tilde p = \left\{ \begin{array}{l} 0
  \quad \mbox{ in R,} \\ e^{-i\omega v} \quad \mbox { in L,}
  \end{array} \right. 
\end{equation}
is the result of `flipping' the mode $p$ over the line $V=0$; it is a
negative-frequency Rindler mode with support in the L wedge. 

We now use the positive-frequency mode $P$ to show that there are
correlations between L and R Rindler modes in the Minkowski
vacuum. Since $P$ has positive frequency with respect to $T$, $a(P) =
(P, \varphi)$ satisfies $a(P) | 0 \rangle = 0$. As 
\begin{equation}
a(P) = a(p) + e^{-\pi \omega/\kappa} a(\tilde p) = a(p) - e^{-\pi
  \omega/\kappa} a^\dagger (\tilde p^*),
\end{equation}
we can rewrite this as 
\begin{equation} \label{rcorr1}
a(p) |0\rangle = e^{-\pi \omega/\kappa} a^\dagger(\tilde p^*) |0\rangle,
\end{equation}
demonstrating the existence of correlations. If we repeat this
argument for a similar positive frequency mode defined starting from
the L wedge, 
\begin{equation}
P' = \left\{ \begin{array}{l} e^{-\pi \omega/\kappa} e^{-i\omega
  \ln(-V)/\kappa} \quad \mbox{ for } V<0 \\ e^{-i\omega \ln(V)/\kappa}
  \quad \mbox { for } V>0, \end{array} \right.
\end{equation}
we can conclude that 
\begin{equation} \label{rcorr2}
a(\tilde p^*) |0\rangle = e^{-\pi \omega/\kappa} a^\dagger(p) |0\rangle.
\end{equation}
These two relations (\ref{rcorr1},\ref{rcorr2}) can be solved by
writing
\begin{equation}
|0 \rangle = \exp \left[ e^{-\pi \omega/\kappa} a^\dagger(\tilde p^*)
 a^\dagger(p) \right] |0_{p \tilde p^*} \rangle,
\end{equation}
where $|0_{p \tilde p^*} \rangle$ is a vacuum state for the modes $p,
\tilde p^*$; that is, a state satisfying 
\begin{equation}
a(\tilde p^*) | 0_{p \tilde p^*} \rangle = a(p) | 0_{p \tilde p^*}
\rangle = 0. 
\end{equation}
If we expand out the exponential, 
\begin{equation}
|0 \rangle = \sum_m e^{-\pi \omega_m/\kappa} |m\rangle_L \times |m \rangle_R,
\end{equation}
where $|m\rangle_{R,L}$ are states with $m$ particles in the mode $p$
($\tilde p^*$) in the R (L) wedge, and $\omega_m$ is the corresponding
energy.  Thus, from the point of view of Rindler observers, the
Minkowski vacuum $|0\rangle$ contains correlated pairs of particles in
the modes $p$, $\tilde p^*$: that is, it is a squeezed state with
respect to the Rindler basis.

This analysis applies to an arbitrary Rindler mode. Formally, we can
solve all the relations by writing
\begin{equation}
|0 \rangle = \prod_i \exp \left[ e^{-\pi \omega_i/\kappa}
 a^\dagger(\tilde p_i^*) a^\dagger(p_i) \right] |0 \rangle_{L} \times
 |0 \rangle_{R},
\end{equation}
for some basis $p_i$ of Rindler modes, where $|0 \rangle_{L,R}$ are
vacuum states wrt to the Rindler modes. However, it should be borne in
mind that this is just a formal expression; the Minkowski vacuum is
not in the same Hilbert space as the Rindler vacuum.

If we consider $\langle 0 | \mathcal{O}_R | 0 \rangle$ for some
operator $\mathcal{O}_R$ which acts only on the R wedge, we trace over
the degrees of freedom in the L wedge, and the correlations between
particles in the L and R wedges will give rise to
thermal behaviour for the expectation value: formally,
\begin{equation}
\langle 0 | \mathcal{O}_R | 0 \rangle = \sum_{m} e^{-2\pi
  \omega_{m}/\kappa} \langle m | \mathcal{O}_R | m \rangle,
\end{equation}
where the sum runs over a complete basis $| m \rangle$ in the Fock
space built on $| 0 \rangle_R$. 

This analysis provides a second, more intuitive point of view on the
thermal nature of the Minkowski vacuum from the Rindler observers'
perspective. Its generalisation will be important in studying Hawking
radiation for a general black hole. The first calculation, however,
avoids reference to any particular mode decomposition, and highlights
an intriguing connection with Euclidean spacetime.

\subsection{Hawking radiation}

We turn now to the thermodynamic properties of black holes. We will
first consider the eternal black hole, and show that there is a
generalisation of the above discussion of Rindler which applies to the
Killing horizon in a stationary black hole spacetime. However, we
should note at the outset that this is not really relevant to black
holes formed by gravitational collapse in asymptotically flat space;
we briefly discuss the argument for Hawking radiation in this truly
dynamical situation at the end of the section.

\begin{figure}[htbp]
\centering
\psfrag{t}{$\tau$}
\psfrag{r}{$r=2M$}
\includegraphics{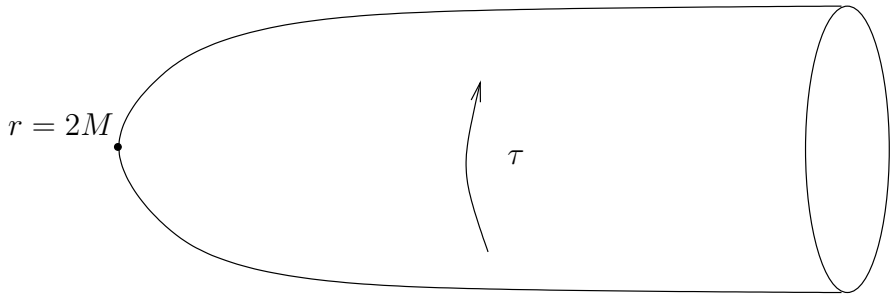}
\caption{The Euclidean black hole geometry, suppressing the two-sphere
  directions $\theta,\phi$. } 
\label{fig10}
\end{figure}

We consider for simplicity the Schwarzschild black hole
\eqref{schw}. As in flat space, we have a time-translation
$\partial_t$ which is timelike in the two wedge regions on the left
and right in figure~\ref{fig1}. If we complexify this time coordinate
by $t \to i \tau$, we obtain the Euclidean metric
\begin{equation} \label{eschw}
ds^2 =  \left( 1 - \frac{2M}{r} \right) d\tau^2 + \left( 1 -
\frac{2M}{r} \right)^{-1} dr^2 + r^2 ( d\theta^2 + \sin^2 \theta
d\phi^2). 
\end{equation}
In this metric, $r=2M$ is an origin in the $r$, $\tau$ plane. The
spacetime is smooth there if $\tau$ is an angular coordinate with
period $\beta = 2\pi/\kappa$ where $\kappa = 1/4M$ is the black hole's
surface gravity. This gives us the `semi-infinite cigar' geometry for
the Euclidean black hole depicted in figure~\ref{fig10}.

If we define a quantum state by taking the Euclidean Green's function
$G(x,y)$ on this spacetime and analytically continuing it to obtain a
two-point function on the Lorentzian black hole spacetime, the
resulting two-point function $G^\beta(t_x - t_y, x,y)$ will satisfy
the KMS condition by the same argument used for the Unruh radiation
above. That is, the state defined by analytic continuation from the
Euclidean black hole, which is referred to as the Hartle-Hawking
state, is a thermal state at temperature $T = \kappa/2\pi$. The
temperature is fixed by the periodicity of $\tau$ required to make the
Euclidean space smooth at the horizon~\cite{gibbonsperry}.

In Rindler space, the analogous state was selected as a preferred
state because it was the Minkowski vacuum. In the Schwarzschild
solution, however, there is no global notion of
time-translation---that is, there is no Killing vector which is
timelike everywhere in figure~\ref{fig1}---so there is no natural
notion of vacuum state.  How then do we choose the Hartle-Hawking
state from the Lorentzian point of view?

The answer is again to demand regularity at the horizon. There is a
theorem due to Kay and Wald~\cite{kaywald} which states that for the
free scalar field on the Schwarzschild geometry, the thermal state at
temperature $T = \kappa/2\pi$ is the unique Hadamard state invariant
under $\partial_t$. Thus, the only thing that can be in equilibrium
with the black hole is thermal radiation at temperature $T$. More
generally,~\cite{kaywald} show that if there exists an invariant
Hadamard state on a stationary spacetime with a bifurcate Killing
horizon, it is a KMS state with temperature $T = \kappa/2\pi$; in
particular, the discussion carries through in exactly the same way for
Schwarzschild-AdS.  Note however that in generalising the black hole
solutions we consider, it can easily happen that {\it no} invariant
Hadamard state exists. For example, none does for Kerr: this is
connected to the superradiance property for Kerr.

Another important point to consider is whether the Hartle-Hawking
state is physically relevant. There are two issues here: first, this
state involves a particular boundary condition at infinity. It
contains an equal flux of incoming and outgoing thermal radiation (as
the state is symmetric under $t \to -t$). As we will argue later, this
is a good approximation to the late-time behaviour of a black hole in
asymptotically AdS space, but not in asymptotically flat
space. Second, we should ask if it describes a {\it stable}
equilibrium. As noted previously, a KMS state is locally dynamically
stable, but the question is now whether the black hole is stable to
small fluctuations. For the Schwarzschild black hole described
above, it is not; since $T = 1/8\pi M$, the specific heat $C_V =
\partial M /\partial T <0$. Thus, if the mass fluctuates downwards, the
temperature rises, and the black hole will radiate more than it
absorbs from the thermal bath, further lowering its mass. So this
equilibrium state for Schwarzschild is unphysical; real black holes
will never reach this equilibrium (even if they are illuminated with
a thermal flux of incoming radiation).

\begin{figure}[htbp]
\centering
\includegraphics[width=0.4\textwidth]{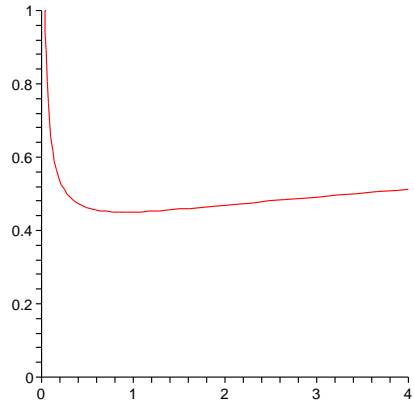}
\caption{The temperature for Schwarzschild-AdS black holes as a
  function of mass, for the case $d=5$. ($T \ell$ is plotted against
  $G M /\ell^2$.)} 
\label{fig11}
\end{figure}

For Schwarzschild-AdS black holes, on the other hand, the temperature
is
\begin{equation} \label{atemp}
T = \frac{(d-1) r_+^2 + (d-3) \ell^2}{4\pi r_+ \ell^2}.
\end{equation}
For $r_+ \ll \ell$, the behaviour is similar to the Schwarzschild
case, but for $r_+^2/\ell^2 = (d-3)/(d-1)$, we reach a minimum
temperature, $T_{min} = \sqrt{(d-1)(d-3)}/2\pi \ell$, and for $r_+ \gg
\ell$,
\begin{equation}
T = \frac{(d-1)}{4\pi \ell^2} r_+ \propto M^{\frac{1}{d-1}}, 
\end{equation}
so the specific heat becomes positive. The temperature is plotted as a
function of mass in figure~\ref{fig11}. Thus for large black hole mass
this becomes a stable equilibrium, and really represents the endstate
of black hole plus radiation. 

\begin{figure}[htbp]
\centering
\psfrag{Im}{$\mathcal{I}^-$}
\psfrag{Ip}{$\mathcal{I}^+$}
\includegraphics[width=0.25\textwidth]{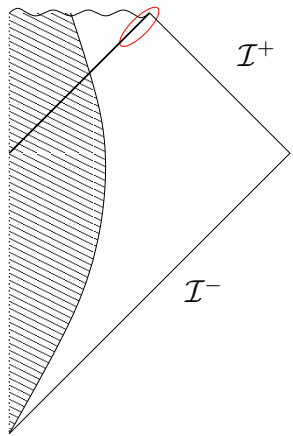}
\caption{The Penrose diagram of the formation of a black hole by
  gravitational collapse in asymptotically flat space. The argument
  for the Hawking effect involves focusing on the circled region.} 
\label{fig9}
\end{figure}

So far, we have discussed quantum fields on the stationary black hole
spacetimes, and analysed the curved space version of the Unruh
effect. However, in the real world, we are interested in the behaviour
of a black hole which forms by the gravitational collapse of a massive
body, as sketched in figure~\ref{fig9}. We will now briefly describe
the argument that such a black hole will emit thermal radiation at
late times---the true Hawking effect.

Once the collapsing matter has crossed the event horizon, the geometry
near the horizon will quickly settle down, approaching an
approximately stationary geometry, which is approximately vacuum on
and outside the event horizon.\footnote{Note that it seems that this
assumption will fail for black holes in asymptotically flat space,
once we take into account the back-reaction of the emitted Hawking
radiation, which causes the black hole to lose mass, balancing the
energy lost to the outgoing radiation. However, this will still be a
good approximation over timescales shorter than the time $t_{br} \sim
(M/M_{pl})^3$ over which Hawking radiation carries away an appreciable
fraction of the mass. For astrophysical black holes, this is a very
long timescale.}  This implies that the geometry of any small
neighbourhood of the future event horizon is approximately flat space,
and the stationary observers outside the black hole are approximately
the Rindler observers in flat space by the argument of
section~\ref{rindler}.

Now comes the crucial assumption: we assume that the quantum state of
the scalar field $\varphi$ on the black hole geometry behaves locally
in the same way as the Minkowski vacuum. That is, freely-falling
observers near the horizon should not see any unusual behaviour in
high-energy processes.\footnote{An interesting question in this
dynamical context is how this free-fall vacuum arises. If we trace the
outgoing Hawking quanta backwards, then due to the redshift noted
previously for the Rindler modes, they will originate in modes of
trans-Planckian frequencies before the collapse. See~\cite{jac} for a
nice review of this trans-Planckian problem.} We are essentially again
assuming the state is regular (Hadamard) at the event horizon. We can
then conclude that the Rindler modes on either side of the horizon are
correlated as in (\ref{rcorr1},\ref{rcorr2}), and the approximately
Rindler observers outside the horizon will see outgoing thermal
radiation at the temperature (with respect to $\partial_t$)
\begin{equation}
T_H = \frac{\hbar \kappa}{2\pi}. 
\end{equation}

However, unlike in our previous discussion, this black hole will not
be in thermal equilibrium with this radiation. This is because the
natural boundary condition on past null infinity $\mathcal I^-$ is
that the state approach the Minkowski vacuum. Thus, there is outgoing
thermal radiation, but no ingoing radiation. The collapse geometry
breaks the $t \to -t$ symmetry, even when we concentrate on the
behaviour long after collapse has taken place. The stationary state on
the eternal black hole which satisfies these boundary conditions is
called the Unruh state (it is not Hadamard on the past horizon). At
late times, a black hole formed from gravitational collapse is
well-approximated by the eternal black hole with the scalar field in
the Unruh state, and not the Hartle-Hawking state. If we take into
account back-reaction, the black hole will slowly reduce in mass,
losing its energy to the outgoing Hawking radiation, so it never
reaches an equilibrium configuration.

The situation is different for an asymptotically AdS black hole. The
asymptotically AdS boundary conditions act as a reflecting box, and
the outgoing Hawking quanta are reflected back towards the black
hole. As a result, at late times the black hole can come to
equilibrium with its Hawking radiation, and for a sufficiently large
black hole, the late-time behaviour is well-approximated by the
eternal black hole with the scalar field in the Hartle-Hawking state.

In this section we have seen that a quantum field in equilibrium with
a black hole must be in a thermal state at temperature $T = \frac{\hbar
\kappa}{2 \pi}$, where $\kappa$ is the black hole's surface
gravity. This is the key result which converts the analogy between
black hole mechanics and thermodynamics in section~\ref{class} into an
equivalence. We can combine this temperature with the first law that
we demonstrated classically in section~\ref{first} to argue that the
black hole has an entropy $S = \frac{A}{4G \hbar}$.

\section{Euclidean entropy calculation}
\label{euclid}

The argument given above provides solid evidence for black hole
thermodynamics. It uses a rigorously defined theory, quantum field
theory on a fixed classical spacetime background. Since the geometry
at the event horizon can be taken to have arbitrarily small curvature
in Planck units, this should be a good approximation to the true
physical description of the situation. In the AdS case, where there is
a stable equilibrium between the black hole and its radiation,
corrections to this description should be small on and outside the
horizon for all times. 

However, having established that the black holes have entropy $S=A/4G
\hbar$, we would like to understand why entropy and area are related
in this way. In this section, I will review the Euclidean path
integral approach, which provides a very general connection between
entropy and area, and discuss the description of black holes in string
theory via the AdS/CFT correspondence using this Euclidean point of
view. The discussion in this section is somewhat briefer and more sketchy
than previously. In particular, have no intention of providing a
review of AdS/CFT; the reader should consult the lectures by Leonardo
Rastelli in this volume, or one of the other reviews
(e.g.,~\cite{agmoo}).

A very general connection between entropy and geometry has been
established in the Euclidean path integral approach to quantum gravity
(see~\cite{hawk,gibb} for reviews). In this approach, the canonical
partition function for the gravitational field is defined by a sum
over all smooth Riemannian geometries which are periodic with period
$\beta = T^{-1}$ in imaginary time,
\begin{equation} \label{pfn}
Z(\beta) = \int d[g] e^{-I[g]}
\end{equation}
where $I[g]$ is the classical action of the geometry. In the
asymptotically flat context, what this means is that the integration
in~\eqref{pfn} includes all asymptotically flat geometries with an
isometry along a compact direction whose proper size at infinity is
$\beta$. In the asymptotically AdS context, if we identify the time
coordinate of the Euclidean black hole solution~\eqref{sads}
periodically, the proper size of this $S^1$ will tend to infinity at
large distances. In this case the relevant quantity to fix is the
asymptotic value of the dimensionless ratio of the size of this $S^1$
to the size of the $S^{d-2}$. In both cases, it is important to note
that we only impose an asymptotic boundary condition on the metric;
the metric in the bulk of the spacetime is allowed to fluctuate.

There are problems with the definition of this Euclidean path
integral: these include the non-renormalisable ultraviolet divergences
of gravity, the indefiniteness of the gravitational action, which is
not even bounded from below, and the fact that the on-shell action
typically diverges for non-compact solutions. I will adopt the view
that the path-integral expression is merely a semi-classical
tool. That is, one should not view the sum over geometries as a
fundamental definition of the theory; instead, we are interested in
seeing what insight we can gain from considering the saddle-point
approximation to this integral, where we approximate
\begin{equation} \label{saddpt}
\ln Z(\beta) \approx -I_s,
\end{equation}
where $I_s$ is the classical action of a Euclidean solution which
satisfies the boundary conditions. There may be more than one such
solution; we consider the dominant contribution, which comes from the
solution of least action. The expectation is that this approximation
should give useful results if the classical solution is weakly curved,
whatever the fundamental quantum theory may be. This expectation is
magnificently borne out in the AdS/CFT context, as we will see a
little later. Since $Z(\beta)$ is the canonical partition function,
$Z(\beta) = e^{-\beta F} = e^{-\beta \langle E \rangle + S}$, we can
evaluate the energy and entropy by the standard formulae
\begin{equation}
\langle E \rangle = -\frac{\partial}{\partial \beta} \ln Z \approx
\frac{\partial}{\partial \beta} I_s, 
\end{equation}
\begin{equation}
S = \beta \langle E \rangle + \ln Z = - \left( \beta
\frac{\partial}{\partial \beta} -1 \right) \ln Z \approx \left( \beta
\frac{\partial}{\partial \beta} -1 \right) I_s. 
\end{equation}

There is an important topological difference between the Euclidean
solutions which do and do not involve black holes. In for example the
Euclidean flat space,
\begin{equation}
ds^2 = d\tau^2 + dr^2 + r^2
d\Omega_{d-2}, 
\end{equation}
the Killing vector $\partial_{\tau}$ is non-vanishing throughout the
spacetime. The radial coordinate $r$ ranges over $r \geq 0$, and the
$S^{d-2}$ shrinks to zero size at $r=0$. We can identify $\tau$
periodically with any period we choose. On the other hand, for the
Euclidean Schwarzschild solution \eqref{eschw}, the Killing vector
$\partial_{\tau}$ vanishes at $r=r_+ = 2M$, which is a fixed point of the
isometry. We must take $r \geq r_+$, and identify $\tau$ periodically
with period $2\pi/\kappa$ to obtain a smooth geometry. The $S^1$
shrinks to zero size at $r=r_+$. This fixed point of $\partial_\tau$
is the Euclidean continuation of the bifurcate Killing horizon in the
Lorentzian black hole solution. 
 
For cases with no black hole, where the circle direction does not
shrink to zero, we can exploit the fact that global time is a Killing
symmetry to write the action as
\begin{equation}
I = \int d^d x  \, \mathcal{L} = \int d\tau \int d^{d-1} x\, \mathcal{L} =
\beta H,
\end{equation}
where $H$ is the Hamiltonian.  Hence, when such a geometry provides
the dominant saddle point, $\ln Z \approx I$ is linear in $\beta$, and
\begin{equation}
S \approx  \left( \beta
\frac{\partial}{\partial \beta} -1 \right) I = 0. 
\end{equation}
That is, there is no classical contribution to the entropy for this
solution, as we would expect. 

\begin{figure}[htbp]
\centering
\psfrag{t}{$\tau$}
\psfrag{r}{$r=r_+$}
\includegraphics{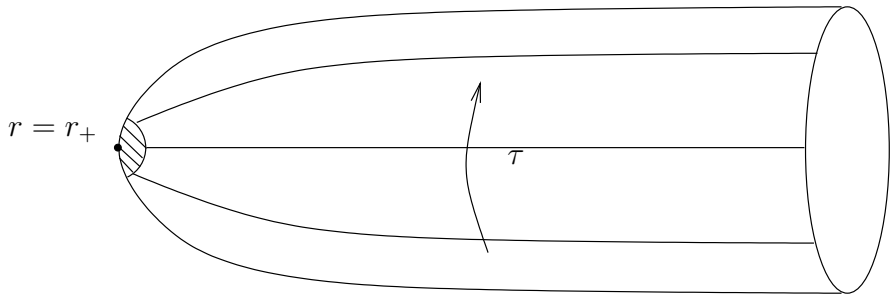}
\caption{Decomposition of the calculation of the bulk action into a
  small region near the horizon and the remainder.} 
\label{fig12}
\end{figure}

For solutions with a black hole, on the other hand, such a foliation
by surfaces of constant time will necessarily break down in the
interior, where the $S^1$ degenerates. Thus, the action will not be
linear in $\beta$. We can split the integration over the spacetime up
in the way shown in figure~\ref{fig12}, into an integral over a small
disc around the horizon at $r=r_+$, and the remaining integration. The
remaining integration will then be linear in $\beta$, as this region
can be foliated with surfaces of constant $t$. 

One might think that the integration over the small disc would vanish
in the limit as we take the size of the disc to zero, since this is a
smooth region of spacetime. However, this is not the case: to be able
to write the integration over the bulk of the spacetime in Hamiltonian
form, we have to be careful about how we break up the integration,
which means we have to keep a boundary term in the action
(see~\cite{banados:disc,hhr}). The (leading-order) gravitational part
of the action for the disc is
\begin{equation}
I_{grav} = \frac{1}{16\pi G} \int_M d^d x \sqrt{-g} R +
\frac{1}{8\pi G} \int_{\partial M} d^{d-1} y \sqrt{-h} K.
\end{equation}
The first term is the usual Einstein-Hilbert term; the second term is
the integral of the trace of the extrinsic curvature over the
boundary, $K = h^{\mu\nu} \nabla_\mu n_\nu$, where $n^\mu$ is the
normal to and $h_{\mu\nu}$ the induced metric on the boundary. The
surface term can also be rewritten as 
\begin{equation}
\int_{\partial M} d^{d-1} y \sqrt{-h} K = - \frac{\partial}{\partial
  n} \int_{\partial M} d^{d-1} y \sqrt{-h}.
\end{equation}
This surface term is necessary to ensure that the variation of the
action vanishes under arbitrary variations of the metric which vanish
on $\partial M$~\cite{gh}.

For a small disc near the horizon, the metric is approximately
\begin{equation}
ds^2 \approx \rho^2 \kappa^2 d\tau^2 + d\rho^2 + r_+^2 d\Omega, 
\end{equation}
so 
\begin{equation}
\int_{r=r_++\epsilon} d^{d-1} y \sqrt{-h} = 2 \pi \epsilon A,
\end{equation}
where $A$ is the area of the horizon, that is, the volume of an
$S^{d-2}$ of radius $r_+$, and
\begin{equation}
\frac{\partial}{\partial
  n} \int_{\partial M} d^{d-1} y \sqrt{-h} = 2 \pi A. 
\end{equation}
Hence, in the limit $\epsilon \to 0$, the small disc around $r=r_+$
makes a contribution
\begin{equation}
I_{disc} = -\frac{1}{4G} A,
\end{equation}
which gives
\begin{equation}
S = \left( \beta \frac{\partial}{\partial \beta} -1 \right) I_s =
\left( \beta \frac{\partial}{\partial \beta} -1 \right) I_{disc} =
\frac{1}{4G} A.
\end{equation}
This calculation provides a direct link between geometry and
entropy. It shows that the non-zero entropy associated to a black hole
is a consequence of the non-trivial topology of the Euclidean
solution. As in the calculation of the temperature in quantum field
theory, regularity of the geometry at the horizon plays a key role in
the derivation. Note that the explicit form of the geometry was not
used in this derivation, just the fact that the geometry is smooth
there. Thus, this derivation explains the universality of the relation
between entropy and area. 

Note also that the explicit form of the action is used here, so the
result depends on the gravitational dynamics (unlike the calculation
of the temperature, which is more kinematical). This Euclidean
calculation has been shown to be equivalent to the Noether charge
construction of the entropy appearing in the first
law~\cite{iywald2}. Hence for a more general action, we would produce
corrections to the above area formula, as in the discussion from the
first law point of view in section~\ref{first}.

This saddle-point calculation of the black hole entropy does not offer
any insight into the nature of the microstates the black hole entropy
is counting. However, in black hole pair creation, a similar instanton
approximation provides evidence that the black hole entropy is really
counting microstates (see~\cite{merev} for a review). There are
processes where a pair of black holes can be created through quantum
tunnelling: for example, a pair of charged black holes can be created
in a sufficiently intense electric or magnetic field. The pair
creation rate can be estimated by instanton methods, using a similar
Euclidean saddle-point. The rate is $\Gamma \sim e^{-I_s}$, where
$I_s$ is the action of the Euclidean instanton. In this context, the
additional contribution to the action associated with the event
horizon corresponds to an enhancement of the pair creation rate by a
factor of $e^{A/4}$ relative to the pair creation rate of elementary
objects of the same mass, just as we would expect if we were summing over
the rate for black holes to be created with in each of $e^{A/4}$
internal states~\cite{hhr}.

To explicitly identify these microstates, we need a concrete
microscopic theory of quantum gravity. I will now briefly describe the
calculation of the entropy in the AdS/CFT correspondence. The
fundamental relation between field theory and spacetime in any of the
AdS/CFT correspondences is that
\begin{equation}
\langle e^{\int \phi_0 \mathcal{O}} \rangle = Z_S(\phi_0).
\end{equation}
The LHS is the generating function of correlation functions in the
field theory, which depends on sources $\phi_0$ coupled to the
operators $\mathcal{O}$ of the theory. This is identified on the RHS
with a string theory partition function in an asymptotically anti-de
Sitter spacetime, with boundary conditions determined by $\phi_0$. The
two most studied examples of this correspondence are: the string
theory on $AdS_5 \times S^5$, with $N$ units of RR five-form flux, is
identified with $\mathcal{N} = 4$, $SU(N)$ Yang-Mills theory in $d=4$,
with coupling $g_{YM}^2 \sim g_s$. Since the curvature scale in $AdS_5
\times S^5$ in string units goes like $g_s N$, the region where a
semiclassical bulk calculation is a good approximation is $g_{YM} \ll
1$, $\lambda = g_{YM}^2 N \gg 1$. The string theory on $AdS_3 \times
S^3 \times X$ for $X = T^4$ or $K3$, with $N_1$ units of electric RR
3-form flux and $N_5$ units of magnetic RR 3-form flux, is identified
a two-dimensional CFT with central charge $c = 6 N_1 N_5$, which is
some deformation of the supersymmetric sigma-model on $X^{N_1
N_5}/S_{N_1 N_5}$. (These correspondences were proposed in~\cite{mads}
and elucidated in~\cite{gkp,witten}. See e.g.~\cite{agmoo} for further
details and references.)
 
We regard the CFT as providing the fundamental definition of the
theory, while on the RHS, we are interested in the semi-classical
approximation, or in perturbative calculation around some
background. Thus, in particular, the canonical partition function in
the CFT, the generating function for thermal correlation functions, is
equal to the string theory version of the Euclidean quantum gravity
partition function~\eqref{pfn}.\footnote{To generate thermal
correlation functions in the CFT, we consider the Euclidean theory on
$S^1 \times S^{d-2}$, as in the previous discussion of the free
scalar. The CFT is strongly interacting, but the argument that
analytic continuation from this Euclidean space defines thermal
correlators still applies.} Thus, the above saddle-point calculation
provides a prediction from gravity for the behaviour of the entropy in
the CFT at strong coupling.

In the CFT, we know precisely what this entropy is counting; it is a
count of the states of a thermal gas of gluons in the field theory. In
the AdS$_5$ case, the gravitational calculation gives a free energy $F
= -\frac{\pi^2}{8} N^2 V T^4$, agreeing up to a numerical factor with
the result obtained at weak coupling by counting the thermal
excitations of the fundamental gauge theory fields, $F = -
\frac{\pi^2}{6} N^2 V T^4$~\cite{gubklebpeet,agmoo}. In the AdS$_3$
case, the spacetime calculation gives $S = 2\pi r_+/4G_{(3)} = 2\pi
\sqrt{c E/3}$, where $E = N_L + N_R = 2N$ is the CFT energy (for a
non-rotating black hole, $N_L = N_R = N$). This result can be
precisely reproduced on the CFT side by using the Cardy formula $S =
2\pi (\sqrt{c N_L/6} + \sqrt{c N_R/6})$ to calculate the
entropy~\cite{strom}. Thus, in this case, we obtain a precise
agreement, including the numerical factor. The central charge of the
CFT can be read off from the asymptotic isometries of the
spacetime~\cite{brownhenn}, so the agreement for AdS$_3$ is largely
independent of the detailed correspondence supplied by AdS/CFT.

Note that these results make no direct reference to the black hole
horizon; the CFT description counts all the states with a given
temperature in asymptotically AdS space. The result is related to the
black hole entropy simply because for high enough temperature the
black hole dominates this ensemble. This has the advantage that it
avoids making any assumption about the spacetime near the horizon---we
do not need to impose any boundary condition at the horizon. The
drawback is that the spacetime interpretation of these microstates,
and their relation to the geometry near the horizon, is obscure.

An interesting aspect of the gravitational calculation is that it
predicts a phase transition as a function of temperature. The
saddle-point which dominates the partition function in the
approximation~\eqref{saddpt} can change as we vary the
temperature. Recall the temperature for a Schwarzschild-AdS black hole
is given by \eqref{atemp}, so for $T<T_{min} = \sqrt{(d-1)(d-3)}/2\pi
\ell$, the only saddle point is thermal AdS. For $T>T_{min} =
\sqrt{(d-1)(d-3)}/2\pi \ell$, there are three saddle-points: thermal
AdS, a small black hole with $r_+ < \sqrt{\frac{(d-3)}{(d-1}} \ell$,
and a large black hole with $r_+ > \sqrt{\frac{(d-3)}{(d-1}} \ell$.
If we define the action by background subtraction, so that $I=0$ for
the Euclidean AdS solution by definition, we find
\begin{equation}
I_{BH} \propto (\ell^2 r_+^{d-2} - r_+^d);
\end{equation}
thus, the action of the black hole changes sign when $r_+ = \ell$.
When $T = (d-2)/2\pi \ell$, so the large black hole has $r_+ = \ell$,
there is a phase transition, the Hawking-Page transition~\cite{hp},
where the dominant contribution to the partition function changes from
thermal AdS to the large black hole. The small black holes, which have
negative specific heat, never make the dominant contribution to the
partition function.

In the AdS$_5$ case, this Hawking-Page transition corresponds to an
expected phase transition in the dual four-dimensional field theory,
the confinement-deconfinement transition, as discussed
in~\cite{wittenthermal}. Thus, the qualitative thermodynamic behaviour
agrees with the CFT expectations. The change in topology of the saddle
plays an important role in the interpretation of the phase transition.
Recently further work has been done studying the phase structure of
the four-dimensional field theory at weak coupling, comparing it to
this behaviour at strong coupling~\cite{aharony}.

In the AdS$_3$ case, we can calculate the elliptic genus of the CFT,
which is a particular supersymmetry-protected partition function,
exactly. In~\cite{farey}, it was shown that the result can be
re-organized in a way which corresponds naturally to the contributions
of different saddle-points in the bulk. Thus, in this case, the
correspondence between the CFT and the sum over geometries in the bulk
can be verified in detail.

There is thus a well-developed relation between the Euclidean black
hole solutions and the thermal ensemble in the field theory. We can
also describe the Lorentzian eternal black hole solution in
AdS/CFT. The essential insight is to take the full Penrose diagram in
figure~\ref{fig4} seriously. The natural dual description is then in
terms of two copies of the CFT, one living on each of the two
asymptotic boundaries. The black hole corresponds to an entangled
state $|\Psi \rangle$ in the product Hilbert space which correlates
the CFT modes living on the two boundaries~\cite{hm,bklt,mald},
\begin{equation}
a_1 |\Psi \rangle = e^{-\pi \omega/\kappa} a^\dagger_2 |\Psi \rangle
\end{equation}
and vice-versa. That is, the whole black hole geometry is described by
having an entanglement between the two CFTs of the same form as the
entanglement between modes of a bulk field on the two sides of a
Killing horizon (cf~\eqref{rcorr1}), even though the two boundaries
cannot communicate through the bulk. This description can be derived
by an analytic continuation from the Euclidean
correspondence~\cite{mald}. The entangled state $|\Psi \rangle$ is a
standard representation of a thermal ensemble: tracing over one copy
of the CFT gives a thermal density matrix in the other copy, so any
observable which makes reference to only one boundary will have
thermal expectation values. This proposed Lorentzian correspondence
has been further developed by a number of
authors~\cite{kos,levi,fhks,kaplan,levi2,brecher,marolf}, probing in
particular the description of the region of spacetime `behind the
horizons'; that is, in the future and past wedges of
figure~\ref{fig4}.

With these recent insights from the AdS/CFT correspondence, we now
have a fairly good picture of the relation between stationary black
holes and thermal equilibrium. There are still interesting unresolved
issues, such as a general discussion of the first law of
thermodynamics for the black rings with `dipole charges', but the most
important open questions concern dynamical situations, such as the
formation and evaporation of a small black hole in AdS. In AdS/CFT,
relatively little progress has been made on these issues: we do not
even have a good understanding of the CFT description of quasi-static
changes in the geometry, let alone a dual description of the formation
of a black hole from gravitational collapse. Addressing such dynamical
questions is the key to addressing many important questions about
black holes, notably the information loss problem~\cite{infoloss}
(see~\cite{pageinfo} for a review). Further progress on these issues
will require a major step forward in our understanding of the AdS/CFT
correspondence, building a dictionary relating bulk and boundary for
dynamical spacetimes and moving away from the essentially Euclidean
picture which has dominated our understanding to date.

\medskip
\centerline{\bf Acknowledgements}
\medskip    

I am grateful to the organisers of the meeting for the invitation to
give these lectures, and I thank the students and other participants
in the meeting for many sharp questions and interesting discussions. I
thank Don Marolf and Veronika Hubeny for carefully reading the draft
version of these notes. This work is supported by the EPSRC.

\bibliographystyle{/home/aplm/dma0sfr/tex_stuff/bibs/utphys}  
 
\bibliography{ubcthermo}   

\providecommand{\href}[2]{#2}\begingroup\raggedright\begin{thebibliography}{10}

\bibitem{jacthermo}
T.~Jacobson, ``Introductory lectures on black hole thermodynamics.'' Given at
  Utrecht U. in 1996; available at
  http://www.glue.umd.edu/\~{}tajac/BHTlectures/lectures.ps.

\bibitem{waldthermo1}
R.~M. Wald, ``Black holes and thermodynamics,''
\href{http://xxx.lanl.gov/abs/gr-qc/9702022}{{\tt gr-qc/9702022}}.

\bibitem{waldthermo2}
R.~M. Wald, ``The thermodynamics of black holes,'' Living Rev. Rel. {\bf 4}
  (2001) 6,
\href{http://xxx.lanl.gov/abs/gr-qc/9912119}{{\tt gr-qc/9912119}}.

\bibitem{jennie}
J.~H. Traschen, ``An introduction to black hole evaporation,''
\href{http://xxx.lanl.gov/abs/gr-qc/0010055}{{\tt gr-qc/0010055}}.

\bibitem{padmanabhan}
T.~Padmanabhan, ``Gravity and the thermodynamics of horizons,'' Phys. Rept.
  {\bf 406} (2005) 49--125,
\href{http://xxx.lanl.gov/abs/gr-qc/0311036}{{\tt gr-qc/0311036}}.

\bibitem{pagethermo}
D.~N. Page, ``Hawking radiation and black hole thermodynamics,''
\href{http://xxx.lanl.gov/abs/hep-th/0409024}{{\tt hep-th/0409024}}.

\bibitem{primer}
R.~Brout, S.~Massar, R.~Parentani, and P.~Spindel, ``A primer for black hole
  quantum physics,'' Phys. Rept. {\bf 260} (1995)
329--454.

\bibitem{fullr}
S.~A. Fulling and S.~N.~M. Ruijsenaars, ``Temperature, periodicity and
  horizons,'' Phys. Repts. {\bf 152} (1987) 135.

\bibitem{wald94}
R.~M. Wald, {\em Quantum field theory in curved space-time and black hole
  thermodynamics}.
\newblock Chicago Univ. Press, Chicago, USA, 1994.

\bibitem{carroll}
S.~M. Carroll, {\em Spacetime and Geometry: An introduction to General
  Relativity}.
\newblock Addison Wesley, San Francisco, 2004.

\bibitem{peet:tasi}
A.~W. Peet, ``{TASI} lectures on black holes in string theory,''
\href{http://xxx.lanl.gov/abs/hep-th/0008241}{{\tt hep-th/0008241}}.

\bibitem{david}
J.~R. David, G.~Mandal, and S.~R. Wadia, ``Microscopic formulation of black
  holes in string theory,'' Phys. Rept. {\bf 369} (2002) 549--686,
\href{http://xxx.lanl.gov/abs/hep-th/0203048}{{\tt hep-th/0203048}}.

\bibitem{hrad1}
S.~W. Hawking, ``Black hole explosions,'' Nature {\bf 248} (1974)
30--31.

\bibitem{hrad2}
S.~W. Hawking, ``Particle creation by black holes,'' Commun. Math. Phys. {\bf
  43} (1975)
199--220.

\bibitem{thooft}
G.~'t~Hooft, ``Dimensional reduction in quantum gravity,''
\href{http://xxx.lanl.gov/abs/gr-qc/9310026}{{\tt gr-qc/9310026}}.

\bibitem{susskind}
L.~Susskind, ``The world as a hologram,'' J. Math. Phys. {\bf 36} (1995)
  6377--6396,
\href{http://xxx.lanl.gov/abs/hep-th/9409089}{{\tt hep-th/9409089}}.

\bibitem{holrev}
R.~Bousso, ``The holographic principle,'' Rev. Mod. Phys. {\bf 74} (2002)
  825--874,
\href{http://xxx.lanl.gov/abs/hep-th/0203101}{{\tt hep-th/0203101}}.

\bibitem{agmoo}
O.~Aharony, S.~S. Gubser, J.~M. Maldacena, H.~Ooguri, and Y.~Oz, ``Large {N}
  field theories, string theory and gravity,'' Phys. Rept. {\bf 323} (2000)
  183--386,
\href{http://xxx.lanl.gov/abs/hep-th/9905111}{{\tt hep-th/9905111}}.

\bibitem{brandenberger}
R.~H. Brandenberger, ``Theory of cosmological perturbations and applications to
  superstring cosmology,''
\href{http://xxx.lanl.gov/abs/hep-th/0501033}{{\tt hep-th/0501033}}.

\bibitem{birdav}
N.~D. Birrell and P.~C.~W. Davies, {\em Quantum fields in curved space}.
\newblock Cambridge Univ. Press, Cambridge, UK, 1982.

\bibitem{haag}
R.~Haag, {\em Local quantum physics}.
\newblock Springer-Verlag, Berlin, 1992.

\bibitem{fullingbook}
S.~A. Fulling, {\em Aspects of quantum field theory in curved space-time},
  vol.~17 of {\em London Mathematical Society student texts}.
\newblock Cambridge Univ. Press, Cambridge, UK, 1989.

\bibitem{dewitt03}
B.~S. DeWitt, {\em The global approach to quantum field theory.}, vol.~114 of
  {\em Int. Ser. Monogr. Phys.}
\newblock Oxford Univ. Press, 2003.
\newblock 2 vols.

\bibitem{dewitt75}
B.~S. Dewitt, ``Quantum field theory in curved space-time,'' Phys. Rept. {\bf
  19} (1975)
295--357.

\bibitem{gibbons79}
G.~W. Gibbons, ``Quantum field theory in curved space-time,'' in {\em General
  Relativity: An Einstein centenary survey}, S.~W. Hawking and W.~Israel, eds.
\newblock Cambridge Univ. Press, Cambridge, UK, 1979.

\bibitem{dewitt79}
B.~S. DeWitt, ``Quantum gravity: The new synthesis,'' in {\em General
  Relativity: An Einstein centenary survey}, S.~W. Hawking and W.~Israel, eds.
\newblock Cambridge Univ. Press, Cambridge, UK, 1979.

\bibitem{ford}
L.~H. Ford, ``Quantum field theory in curved spacetime,''
\href{http://xxx.lanl.gov/abs/gr-qc/9707062}{{\tt gr-qc/9707062}}.

\bibitem{jac}
T.~Jacobson, ``Introduction to quantum fields in curved spacetime and the
  {H}awking effect,''
\href{http://xxx.lanl.gov/abs/gr-qc/0308048}{{\tt gr-qc/0308048}}.

\bibitem{penconf}
R.~Penrose, ``Conformal treatment of infinity,'' in {\em Relativity, groups and
  topology}, C.~M. de~Witt and B.~de~Witt, eds.
\newblock Gordon \& Breach, New York, 1964.

\bibitem{pendiag}
R.~Penrose, ``Zero rest mass fields including gravitation: Asymptotic
  behavior,'' Proc. Roy. Soc. Lond. {\bf A284} (1965)
159.

\bibitem{adspen}
L.~Fidkowski, V.~Hubeny, M.~Kleban, and S.~Shenker, ``The black hole
  singularity in ads/cft,'' JHEP {\bf 02} (2004) 014,
\href{http://xxx.lanl.gov/abs/hep-th/0306170}{{\tt hep-th/0306170}}.

\bibitem{vacnohair}
W.~Israel, ``Event horizons in static vacuum space-times,'' Phys. Rev. {\bf
  164} (1967)
1776--1779.

\bibitem{nohairhigher}
G.~W. Gibbons, D.~Ida, and T.~Shiromizu, ``Uniqueness and non-uniqueness of
  static black holes in higher dimensions,'' Phys. Rev. Lett. {\bf 89} (2002)
  041101,
\href{http://xxx.lanl.gov/abs/hep-th/0206049}{{\tt hep-th/0206049}}.

\bibitem{heusler}
M.~Heusler, ``Stationary black holes: Uniqueness and beyond,'' Living Rev. Rel.
  {\bf 1} (1998)
6.

\bibitem{er}
R.~Emparan and H.~S. Reall, ``A rotating black ring in five dimensions,'' Phys.
  Rev. Lett. {\bf 88} (2002) 101101,
\href{http://xxx.lanl.gov/abs/hep-th/0110260}{{\tt hep-th/0110260}}.

\bibitem{hawkellis}
S.~W. Hawking and G.~F.~R. Ellis, {\em The large scale structure of
  space-time}.
\newblock Cambridge Univ. Press, Cambridge, UK, 1973.

\bibitem{carter}
B.~Carter, ``Black hole equilibrium states,'' in {\em Black Holes}, C.~DeWitt
  and B.~S. DeWitt, eds.
\newblock Gordon \& Breach, New York, 1973.

\bibitem{town}
P.~K. Townsend, ``Black holes,''
\href{http://xxx.lanl.gov/abs/gr-qc/9707012}{{\tt gr-qc/9707012}}.

\bibitem{harea}
S.~W. Hawking, ``Gravitational radiation from colliding black holes,'' Phys.
  Rev. Lett. {\bf 26} (1971)
1344--1346.

\bibitem{bek}
J.~D. Bekenstein, ``Black holes and entropy,'' Phys. Rev. D {\bf 7} (1973)
2333--2346.

\bibitem{bch}
J.~Bardeen, B.~Carter, and S.~W. Hawking, ``The four laws of black hole
  mechanics,'' Commun. Math. Phys. {\bf 31} (1973) 161.

\bibitem{wiseman}
T.~Wiseman, ``Static axisymmetric vacuum solutions and non-uniform black
  strings,'' Class. Quant. Grav. {\bf 20} (2003) 1137--1176,
\href{http://xxx.lanl.gov/abs/hep-th/0209051}{{\tt hep-th/0209051}}.

\bibitem{infnon}
R.~Emparan, ``Rotating circular strings, and infinite non-uniqueness of black
  rings,'' JHEP {\bf 03} (2004) 064,
\href{http://xxx.lanl.gov/abs/hep-th/0402149}{{\tt hep-th/0402149}}.

\bibitem{isol}
A.~Ashtekar {\em et.~al.}, ``Isolated horizons and their applications,'' Phys.
  Rev. Lett. {\bf 85} (2000) 3564--3567,
\href{http://xxx.lanl.gov/abs/gr-qc/0006006}{{\tt gr-qc/0006006}}.

\bibitem{wald}
R.~M. Wald, ``Black hole entropy in the {N}oether charge,'' Phys. Rev. D {\bf
  48} (1993) 3427--3431,
\href{http://xxx.lanl.gov/abs/gr-qc/9307038}{{\tt gr-qc/9307038}}.

\bibitem{iywald}
V.~Iyer and R.~M. Wald, ``Some properties of {N}oether charge and a proposal
  for dynamical black hole entropy,'' Phys. Rev. {\bf D50} (1994) 846--864,
\href{http://xxx.lanl.gov/abs/gr-qc/9403028}{{\tt gr-qc/9403028}}.

\bibitem{sudwald}
D.~Sudarsky and R.~M. Wald, ``Extrema of mass, stationarity, and staticity, and
  solutions to the {Einstein Yang-Mills} equations,'' Phys. Rev. D {\bf 46}
  (1992)
1453--1474.

\bibitem{jacmy}
T.~Jacobson and R.~C. Myers, ``Black hole entropy and higher curvature
  interactions,'' Phys. Rev. Lett. {\bf 70} (1993) 3684--3687,
\href{http://xxx.lanl.gov/abs/hep-th/9305016}{{\tt hep-th/9305016}}.

\bibitem{dewit1}
G.~Lopes~Cardoso, B.~de~Wit, and T.~Mohaupt, ``Corrections to macroscopic
  supersymmetric black-hole entropy,'' Phys. Lett. {\bf B451} (1999) 309--316,
\href{http://xxx.lanl.gov/abs/hep-th/9812082}{{\tt hep-th/9812082}}.

\bibitem{dewit2}
G.~Lopes~Cardoso, B.~de~Wit, and T.~Mohaupt, ``Deviations from the area law for
  supersymmetric black holes,'' Fortsch. Phys. {\bf 48} (2000) 49--64,
\href{http://xxx.lanl.gov/abs/hep-th/9904005}{{\tt hep-th/9904005}}.

\bibitem{dewit3}
G.~Lopes~Cardoso, B.~de~Wit, and T.~Mohaupt, ``Macroscopic entropy formulae and
  non-holomorphic corrections for supersymmetric black holes,'' Nucl. Phys.
  {\bf B567} (2000) 87--110,
\href{http://xxx.lanl.gov/abs/hep-th/9906094}{{\tt hep-th/9906094}}.

\bibitem{msw}
J.~M. Maldacena, A.~Strominger, and E.~Witten, ``Black hole entropy in
  {M}-theory,'' JHEP {\bf 12} (1997) 002,
\href{http://xxx.lanl.gov/abs/hep-th/9711053}{{\tt hep-th/9711053}}.

\bibitem{vafa}
C.~Vafa, ``Black holes and {C}alabi-{Y}au threefolds,'' Adv. Theor. Math. Phys.
  {\bf 2} (1998) 207--218,
\href{http://xxx.lanl.gov/abs/hep-th/9711067}{{\tt hep-th/9711067}}.

\bibitem{osv}
H.~Ooguri, A.~Strominger, and C.~Vafa, ``Black hole attractors and the
  topological string,''
\href{http://xxx.lanl.gov/abs/hep-th/0405146}{{\tt hep-th/0405146}}.

\bibitem{jkm}
T.~Jacobson, G.~Kang, and R.~C. Myers, ``Increase of black hole entropy in
  higher curvature gravity,'' Phys. Rev. D {\bf 52} (1995) 3518--3528,
\href{http://xxx.lanl.gov/abs/gr-qc/9503020}{{\tt gr-qc/9503020}}.

\bibitem{minkquot}
H.~Minkowski, ``Space and time,'' 1908.
\newblock Address at the 80th Assembly of German Natural Scientists and
  Physicians.

\bibitem{peierls}
R.~E. Peierls, ``The commutation laws of relativistic field theory,'' Proc.
  Roy. Soc. (London) {\bf A214} (1952) 143.

\bibitem{kaywald}
B.~S. Kay and R.~M. Wald, ``Theorems on the uniqueness and thermal properties
  of stationary, nonsingular, quasifree states on space-times with a bifurcate
  {K}illing horizon,'' Phys. Rept. {\bf 207} (1991)
49--136.

\bibitem{kubo}
R.~Kubo, ``Statistical mechanical theory of irreversible processes. 1.
  {G}eneral theory and simple applications in magnetic and conduction
  problems,'' J. Phys. Soc. Jap. {\bf 12} (1957)
570--586.

\bibitem{ms}
P.~C. Martin and J.~S. Schwinger, ``Theory of many particle systems. {I},''
  Phys. Rev. {\bf 115} (1959)
1342--1373.

\bibitem{sewellbook}
G.~L. Sewell, {\em Quantum theory of collective phenomena}.
\newblock Oxford Univ. Press, Oxford, UK, 1986.

\bibitem{unruh}
W.~G. Unruh, ``Notes on black hole evaporation,'' Phys. Rev. D {\bf 14} (1976)
870.

\bibitem{bw1}
J.~J. Bisognano and E.~H. Wichmann, ``On the duality condition for a
  {H}ermitian scalar field,'' J. Math. Phys. {\bf 16} (1975)
985--1007.

\bibitem{bw2}
J.~J. Bisognano and E.~H. Wichmann, ``On the duality condition for quantum
  fields,'' J. Math. Phys. {\bf 17} (1976)
303--321.

\bibitem{sewell}
G.~L. Sewell, ``Quantum fields on manifolds: {PCT} and gravitationally indiced
  thermal states,'' Ann. Phys. {\bf 141} (1982) 201.

\bibitem{gibbonsperry}
G.~W. Gibbons and M.~J. Perry, ``Black holes and thermal {G}reen's functions,''
  Proc. Roy. Soc. Lond. {\bf A358} (1978)
467--494.

\bibitem{hawk}
S.~W. Hawking, ``The path-integral approach to quantum gravity,'' in {\em
  General Relativity: An Einstein Centenary Survey}, S.~W. Hawking and
  W.~Israel, eds., pp.~746--789.
\newblock Cambridge Univ. Press, Cambridge, UK, 1979.

\bibitem{gibb}
G.~W. Gibbons, ``Euclidean quantum gravity: the view from 2002,'' in {\em The
  Future of Theoretical Physics and Cosmology}, G.~W. Gibbons, E.~P.~S.
  Shellard, and S.~J. Rankin, eds.
\newblock Cambridge Univ. Press, Cambridge, UK, 2003.

\bibitem{banados:disc}
M.~Banados, C.~Teitelboim, and J.~Zanelli, ``Black hole entropy and the
  dimensional continuation of the gauss-bonnet theorem,'' Phys. Rev. Lett. {\bf
  72} (1994) 957--960,
\href{http://xxx.lanl.gov/abs/gr-qc/9309026}{{\tt gr-qc/9309026}}.

\bibitem{hhr}
S.~W. Hawking, G.~T. Horowitz, and S.~F. Ross, ``Entropy, area, and black hole
  pairs,'' Phys. Rev. D {\bf 51} (1995) 4302--4314,
\href{http://xxx.lanl.gov/abs/gr-qc/9409013}{{\tt gr-qc/9409013}}.

\bibitem{gh}
G.~W. Gibbons and S.~W. Hawking, ``Action integrals and partition functions in
  quantum gravity,'' Phys. Rev. D {\bf 15} (1977)
2752--2756.

\bibitem{iywald2}
V.~Iyer and R.~M. Wald, ``A comparison of {N}oether charge and {E}uclidean
  methods for computing the entropy of stationary black holes,'' Phys. Rev.
  {\bf D52} (1995) 4430--4439,
\href{http://xxx.lanl.gov/abs/gr-qc/9503052}{{\tt gr-qc/9503052}}.

\bibitem{merev}
S.~F. Ross, ``Black hole pair creation,'' in {\em The Future of Theoretical
  Physics and Cosmology}, G.~W. Gibbons, E.~P.~S. Shellard, and S.~J. Rankin,
  eds.
\newblock Cambridge Univ. Press, Cambridge, UK, 2003.

\bibitem{mads}
J.~M. Maldacena, ``The large {N} limit of superconformal field theories and
  supergravity,'' Adv. Theor. Math. Phys. {\bf 2} (1998) 231--252,
\href{http://xxx.lanl.gov/abs/hep-th/9711200}{{\tt hep-th/9711200}}.

\bibitem{gkp}
S.~S. Gubser, I.~R. Klebanov, and A.~M. Polyakov, ``Gauge theory correlators
  from non-critical string theory,'' Phys. Lett. {\bf B428} (1998) 105--114,
\href{http://xxx.lanl.gov/abs/hep-th/9802109}{{\tt hep-th/9802109}}.

\bibitem{witten}
E.~Witten, ``Anti-de {S}itter space and holography,'' Adv. Theor. Math. Phys.
  {\bf 2} (1998) 253--291,
\href{http://xxx.lanl.gov/abs/hep-th/9802150}{{\tt hep-th/9802150}}.

\bibitem{gubklebpeet}
S.~S. Gubser, I.~R. Klebanov, and A.~W. Peet, ``Entropy and temperature of
  black 3-branes,'' Phys. Rev. D {\bf 54} (1996) 3915--3919,
\href{http://xxx.lanl.gov/abs/hep-th/9602135}{{\tt hep-th/9602135}}.

\bibitem{strom}
A.~Strominger, ``Black hole entropy from near-horizon microstates,'' JHEP {\bf
  02} (1998) 009,
\href{http://xxx.lanl.gov/abs/hep-th/9712251}{{\tt hep-th/9712251}}.

\bibitem{brownhenn}
J.~D. Brown and M.~Henneaux, ``Central charges in the canonical realization of
  asymptotic symmetries: An example from three-dimensional gravity,'' Commun.
  Math. Phys. {\bf 104} (1986)
207--226.

\bibitem{hp}
S.~W. Hawking and D.~N. Page, ``Thermodynamics of black holes in anti-de
  {S}itter space,'' Commun. Math. Phys. {\bf 87} (1983)
577.

\bibitem{wittenthermal}
E.~Witten, ``Anti-de {S}itter space, thermal phase transition, and confinement
  in gauge theories,'' Adv. Theor. Math. Phys. {\bf 2} (1998) 505--532,
\href{http://xxx.lanl.gov/abs/hep-th/9803131}{{\tt hep-th/9803131}}.

\bibitem{aharony}
O.~Aharony, J.~Marsano, S.~Minwalla, K.~Papadodimas, and M.~Van~Raamsdonk,
  ``The {H}agedorn/deconfinement phase transition in weakly coupled large {$N$}
  gauge theories,''
\href{http://xxx.lanl.gov/abs/hep-th/0310285}{{\tt hep-th/0310285}}.

\bibitem{farey}
R.~Dijkgraaf, J.~M. Maldacena, G.~W. Moore, and E.~Verlinde, ``A black hole
  farey tail,''
\href{http://xxx.lanl.gov/abs/hep-th/0005003}{{\tt hep-th/0005003}}.

\bibitem{hm}
G.~T. Horowitz and D.~Marolf, ``A new approach to string cosmology,'' JHEP {\bf
  07} (1998) 014,
\href{http://xxx.lanl.gov/abs/hep-th/9805207}{{\tt hep-th/9805207}}.

\bibitem{bklt}
V.~Balasubramanian, P.~Kraus, A.~E. Lawrence, and S.~P. Trivedi, ``Holographic
  probes of anti-de {S}itter space-times,'' Phys. Rev. D {\bf 59} (1999)
  104021,
\href{http://xxx.lanl.gov/abs/hep-th/9808017}{{\tt hep-th/9808017}}.

\bibitem{mald}
J.~M. Maldacena, ``Eternal black holes in anti-de-{S}itter,'' JHEP {\bf 04}
  (2003) 021,
\href{http://xxx.lanl.gov/abs/hep-th/0106112}{{\tt hep-th/0106112}}.

\bibitem{kos}
P.~Kraus, H.~Ooguri, and S.~Shenker, ``Inside the horizon with {AdS/CFT},''
  Phys. Rev. D {\bf 67} (2003) 124022,
\href{http://xxx.lanl.gov/abs/hep-th/0212277}{{\tt hep-th/0212277}}.

\bibitem{levi}
T.~S. Levi and S.~F. Ross, ``Holography beyond the horizon and cosmic
  censorship,'' Phys. Rev. D {\bf 68} (2003) 044005,
\href{http://xxx.lanl.gov/abs/hep-th/0304150}{{\tt hep-th/0304150}}.

\bibitem{fhks}
L.~Fidkowski, V.~Hubeny, M.~Kleban, and S.~Shenker, ``The black hole
  singularity in {AdS/CFT},'' JHEP {\bf 02} (2004) 014,
\href{http://xxx.lanl.gov/abs/hep-th/0306170}{{\tt hep-th/0306170}}.

\bibitem{kaplan}
J.~Kaplan, ``Extracting data from behind horizons with the {AdS/CFT}
  correspondence,''
\href{http://xxx.lanl.gov/abs/hep-th/0402066}{{\tt hep-th/0402066}}.

\bibitem{levi2}
V.~Balasubramanian and T.~S. Levi, ``Beyond the veil: Inner horizon instability
  and holography,'' Phys. Rev. D {\bf 70} (2004) 106005,
\href{http://xxx.lanl.gov/abs/hep-th/0405048}{{\tt hep-th/0405048}}.

\bibitem{brecher}
D.~Brecher, J.~He, and M.~Rozali, ``On charged black holes in anti-de {S}itter
  space,''
\href{http://xxx.lanl.gov/abs/hep-th/0410214}{{\tt hep-th/0410214}}.

\bibitem{marolf}
D.~Marolf, ``States and boundary terms: Subtleties of {L}orentzian {AdS/CFT},''
\href{http://xxx.lanl.gov/abs/hep-th/0412032}{{\tt hep-th/0412032}}.

\bibitem{infoloss}
S.~W. Hawking, ``Breakdown of predictability in gravitational collapse,'' Phys.
  Rev. D {\bf 14} (1976)
2460--2473.

\bibitem{pageinfo}
D.~N. Page, ``Black hole information,''
\href{http://xxx.lanl.gov/abs/hep-th/9305040}{{\tt hep-th/9305040}}.

\end{thebibliography}\endgroup

\end{document}